\documentclass{aa}

\usepackage{natbib}
\bibpunct{(}{)}{;}{a}{}{,} 

\usepackage{latexsym,exscale,amssymb,amsmath}
\usepackage{eurosym}
\usepackage{color}
\usepackage{graphicx}
\usepackage[percent]{overpic}
\usepackage{hyperref}

\title{\bf [C\,{\sc ii}] $158\,\mu\mathrm{m}$ line emission from Orion A.\\ I. A template for extragalactic studies?}
\titlerunning{[C\,{\sc ii}] $158\,\mu\mathrm{m}$ line emission from Orion A I}
\author{C. H. M. Pabst\inst{\ref{inst1}} \and A. Hacar\inst{\ref{inst1},\ref{inst11}} \and J. R. Goicoechea\inst{\ref{inst2}} \and D. Teyssier\inst{\ref{inst3}} \and O. Bern\'{e}\inst{\ref{inst4}} \and M. G. Wolfire\inst{\ref{inst5}} \and R. D. Higgins\inst{\ref{inst6}} \and E. T. Chambers\inst{\ref{inst7}} \and S. Kabanovic\inst{\ref{inst6}} \and R. G\"{u}sten\inst{\ref{inst8}} \and J. Stutzki\inst{\ref{inst6}} \and C. Kramer\inst{\ref{inst9}} \and A. G. G. M. Tielens\inst{\ref{inst1},\ref{inst5}} }

\institute{Leiden Observatory, Leiden University, Niels Bohrweg 2, 2333 CA Leiden, Netherlands\label{inst1}; \href{mailto:pabst@strw.leidenuniv.nl}{\texttt{pabst@strw.leidenuniv.nl}}
\and University of Vienna, Department of Astrophysics, T\"{u}rkenschanzstrasse 17, 1180 Vienna, Austria\label{inst11}
\and Instituto de F\'{\i}sica Fundamental, CSIC, Calle Serrano 121-123, 28006 Madrid, Spain \label{inst2} 
\and Telespazio Vega UK Ltd. for ESA/ESAC, Urbanizacion Villafranca del Castillo, 28691 Madrid, Spain \label{inst3}
\and IRAP, Universit\'{e} de Toulouse, CNRS, CNES, UPS, 9 Av. colonel Roche, 31028 Toulouse Cedex 4, France \label{inst4}
\and Department of Astronomy, University of Maryland, College Park, MD 20742, USA\label{inst5}
\and I. Physikalisches Institut der Universit\"{a}t zu K\"{o}ln, Z\"{u}lpicher Strasse 77, 50937 K\"{o}ln, Germany\label{inst6}
\and USRA/SOFIA, NASA Ames Research Center, Mail Stop 232-12, Building N232, P.O. Box 1, Moffett Field, CA 94035-0001, USA\label{inst7}
\and Max-Planck-Institut f\"{u}r Radioastronomie, Auf dem H\"{u}gel 69, 53121 Bonn, Germany\label{inst8}
\and Institut de Radioastronomie Millim\'etrique, 300 rue de la Piscine, 38406 Saint Martin d'H\`{e}res, France\label{inst9}
}
\date{Received 13 March 2021, Accepted ---}

\abstract{The [C\,{\sc ii}] $158\,\mu\mathrm{m}$ fine-structure line is one of the dominant coolants of the neutral interstellar medium. It is hence one of the brightest far-infrared emission lines and can be observed not only in star-forming regions throughout the Galaxy, but also in the diffuse interstellar medium and in distant galaxies. [C\,{\sc ii}] line emission has been suggested to be a powerful tracer of star-formation.}{We aim to understand the origin of [C\,{\sc ii}] emission and its relation to other tracers of interstellar gas and dust. This includes a study of the heating efficiency of interstellar gas as traced by the [C\,{\sc ii}] line to test models of gas heating.}{We make use of a one-square-degree map of velocity-resolved [C\,{\sc ii}] line emission towards the Orion Nebula complex, including M43 and NGC 1977. We employ {\it Herschel} far-infrared photometric images to determine dust properties. Moreover, we compare with H$\alpha$ emission from the ionized gas, {\it Spitzer} mid-infrared photometry to trace hot dust and large polycyclic aromatic hydrocarbons (PAHs), and velocity-resolved IRAM 30m CO(2-1) observations of the molecular gas.}{The [C\,{\sc ii}] intensity is tightly correlated with PAH emission in the IRAC $8\,\mu\mathrm{m}$ band and far-infrared emission from warm dust. However, the [C\,{\sc ii}] intensity depends less than linear on the $8\,\mu\mathrm{m}$ and far-infrared intensity, while $8\,\mu\mathrm{m}$ and far-infrared intensities are approximately linearly correlated. The correlation between [C\,{\sc ii}] and CO(2-1) does not show a clear trend and is affected by the detailed geometry of the region. We find particularly low [C\,{\sc ii}]-over-FIR intensity ratios towards large columns of (warm and cold) dust, which suggest the interpretation of the ``[C\,{\sc ii}] deficit'' in terms of a ``FIR excess''. A slight decrease in the FIR line-over-continuum intensity ratio can be attributed to a decreased heating efficiency of the gas. Furthermore, we find that, at the mapped spatial scales, predictions of the star-formation rate from [C\,{\sc ii}] emission, like most other tracers, underestimate the star-formation rate calculated from YSO counts in the Orion Nebula complex by an order of magnitude.}{[C\,{\sc ii}] emission from the Orion Nebula complex arises dominantly in the cloud surfaces, many viewed in edge-on geometry. Most of the [C\,{\sc ii}] emission stems from the extended fainter outskirts of the irradiated regions, while the [C\,{\sc ii}] intensity is deficient with respect to the total far-infrared intensity in the brightest regions. [C\,{\sc ii}] emission from extended faint cloud surfaces may contribute significantly to the total [C\,{\sc ii}] emission on galactic scales.}


\begin{document}

\maketitle

\section{Introduction}

The [C\,{\sc ii}] $158\,\mu\mathrm{m}$ fine-structure line dominates the far-infrared (FIR) spectrum of the Milky Way, carrying up to 1\%\ of the total luminosity \citep{Bennett1994}. Given the small energy separation of the two fine structure-levels ($\Delta E=91.2\,\mathrm{K}$), low critical density \citep[several 10$^3$\,cm$^{-3}$ for collisions with hydrogen atoms; e.g.,][]{Goldsmith2012}, and the high abundance of carbon, this transition is the dominant cooling line of diffuse atomic gas \citep{DalgarnoMcCray1972}, the so-called cold neutral medium phase of the interstellar medium (ISM). In the surfaces of molecular clouds, the CO molecule is photodissociated and carbon is singly ionized (ionization potential of 11.3\,eV) by the interstellar radiation field. This so-called CO-dark (atomic and) molecular gas \citep{Grenier2005,Wolfire2010} also radiates substantially in the [C\,{\sc ii}] line \citep[e.g.,][]{Langer2010}. Also in regions of massive star formation, the strong UV field photodissociates molecules, ionizes atoms, and heats the gas. These dense photodissociation regions (PDRs) emit bright atomic fine-structure lines. In particular, the [C\,{\sc ii}] 158\,$\mu$m line and the [O\,{\sc i}] 63\,$\mu$m and 145\,$\mu$m lines (at high gas densities) are the main gas coolants \citep{HollenbachTielens1999}. Finally, the [C\,{\sc ii}] line can also arise from low-density regions of ionized gas, such as the warm ionized medium \citep{Heiles1994}. Observational studies have demonstrated that each of these ISM components contributes to the [C\,{\sc ii}] emission of the Milky Way \citep[e.g.,][]{Stacey1991,Shibai1991,Bennett1994,Bock1993,Abdullah2020}. In particular, the pencil-beam Herschel/HIFI GOTC+ [C\,{\sc ii}] survey suggests that much of the observed [C\,{\sc ii}] emission in our Galaxy emerges from dense PDRs ($\sim$\,47\%), with smaller contributions from CO-dark molecular gas ($\sim$\,28\,\%), cold atomic gas ($\sim$\,21\,\%), and ionized gas \citep[$\sim$\,4\,\%;][]{Pineda2013}.

Irrespective of its origin, the [C\,{\sc ii}] emission is linked to the presence of stellar far-ultraviolet photons \mbox{(FUV; $E<13.6\,\mathrm{eV}$)}. As FUV photons are tied to the presence of massive O and B stars that have short lifetimes, the [C\,{\sc ii}]\,158\,$\mu$m line is also a star formation rate indicator. Indeed, ISO, Herschel and SOFIA observations have demonstrated the good correlation between the [C\,{\sc ii}] luminosity and the star-formation rate (SFR) in the Milky Way and in regions of massive star formation in other galaxies \citep[e.g.,][]{Kramer2013, Kramer2020,Pineda2014,Pineda2018,Herrera-Camus2015,Herrera-Camus2018b,DeLooze2011}. With ALMA and NOEMA, ground-based  observations of the [C\,{\sc ii}] 158\,$\mu$m line in high redshift galaxies has come into reach and such data is routinely used to infer star formation rates \citep[e.g.,][]{Walter2012,Venemans2012,Knudsen2016,Bischetti2018,Khusanova2020} based upon validations of this relationship in the nearby Universe \citep{Herrera-Camus2018b,DeLooze2011}. However, it is well-understood that the intensity of the [C\,{\sc ii}] line depends on the local physical conditions \citep{HollenbachTielens1999}. Observationally, the presence of the so-called ``[C\,{\sc ii}]-deficit'' -- a decreased ratio of [C\,{\sc ii}] 158 $\mu$m luminosity to FIR dust continuum with increasing dust color temperature and also with FIR luminosity -- is well established \citep{Malhotra2001,Diaz-Santos2013,Magdis2014,Smith2017}. This deficit is particularly pronounced in (local) ultraluminous infrared galaxies (ULIRGs), very dusty galaxies characterized by vigorous embedded star formation \citep[e.g.,][]{Luhman2003,Abel2009,Gracia2011}. This deficit, however, does not necessarily hold in the early Universe at high redshift \citep[e.g.,][]{Stacey2010,Brisbin2015,Capak2015}. Some studies have indicated that not only [C\,{\sc ii}] emission is deficient in some sources, but other FIR cooling lines ([O\,{\sc i}], [O\,{\sc iii}], [N\,{\sc ii}), as well \citep[e.g.,][]{Gracia2011,Herrera-Camus2018b}. These deficits must be linked to the global ISM properties and star-formation characteristics in these galaxies.

As these FIR cooling lines and FIR dust continuum emission are often the brightest signature from the ISM of distant star-forming galaxies, their luminosities should allow us to determine galaxy-averaged gas physical conditions, SFRs, and ultimately distinguish the dominant star-forming modes and galaxy evolutionary stages (e.g., starbursts, mergers). Before exploiting such a powerful diagnostic tool box of star-formation across cosmic time, it is mandatory to fully understand the origin of the FIR fine-structure line emission, its link with the SFR, and how does it relate to other proxies of the SFR (dust, H$\alpha$, PAH, and molecular gas emission). Here we try to answer these questions with square-degree \mbox{[C\,{\sc ii}] 158\,$\mu$m} mapping observations of the prototypical star-forming cloud in the disk of the Milky Way, Orion A \citep{Pabst2019}. Our study is relevant in the extragalactic context because, thanks to Herschel and now much more efficiently with SOFIA, we have access to velocity-resolved wide-field (spatial scales of several parsec) maps of the \mbox{[C\,{\sc ii}] 158\,$\mu$m} emission over nearby, spatially-resolved regions of massive star formation \citep{Goicoechea2015,Pabst2019,Pabst2020,Schneider2020}. This means that any variation of the surface brightness spatial distribution of these observables can be directly linked to the characteristics of each line-of-sight, its local physical and \mbox{FUV-illumination} conditions, and its SFR surface density. Mapping large spatial scales of star-forming regions in the Milky Way is mandatory to properly compare with the extragalactic emission. As we show here, the \mbox{[C\,{\sc ii}]-to-infrared} dust intensity ratio over the mapped areas in Orion A varies by about two orders of magnitude, showing the same range of intensity ratios displayed by galaxies hosting very different physical conditions and SFRs \citep[e.g.,][]{Stacey2010}. This similarity suggests that Orion A is a good local template to understand the origin of the \mbox{[C\,{\sc ii}] 158\,$\mu$m} emission and its connection to the SFR in galaxies \citep[see also the discussion by][]{Goicoechea2015,Rybak2020b}. In order to expand our study and include regions of  low \mbox{[C\,{\sc ii}] 158\,$\mu$m} surface brightness into the analysis (i.e. low FUV illumination), and to investigate the role of different environment/stellar content (at solar metallicties), we also compare with the [C\,{\sc ii}] emission seen around the Horsehead Nebula \citep[in Orion B, about $3^{\circ}$ north of Orion A;][]{Pabst2017,Bally2018} and also with the [C\,{\sc ii}] emission imaged over the entire nearby spiral galaxy M51 \citep{Pineda2018}, both mapped by SOFIA.

This study is organized into a series of two papers. In this paper (Paper I), we examine the relationship between the [C\,{\sc ii}] emission and other observational tracers of star formation, as well as the implications for the [C\,{\sc ii}] deficit, using the square-degree observations of Orion carried out with SOFIA/upGREAT \citep[for details see][]{Pabst2019,Pabst2020}. In a second paper (Paper II), we investigate the correlations in more detail, relate them to small-scale emission features and PDR physics, and compare with detailed PDR models. This paper is organized as follows: In Section 2, we will briefly describe the observations used in this study. Section 3 discusses the global morphology of the [C\,{\sc ii}] line emission and presents the correlations we find of the [C\,{\sc ii}] emission with other tracers of gas and dust in the Orion Nebula complex. In Section 4, we discuss the implications of the results of Section 3 and the relation between [C\,{\sc ii}] emission and the star-formation rate. We summarize our results in Section 5.

\section{Observations}

\subsection{[C\,{\sc ii}] observations}

We use the same velocity-resolved [C\,{\sc ii}] line observations as \cite{Pabst2019,Pabst2020}. Details of the observing strategy and data reduction are provided by \cite{Higgins2021}. The [C\,{\sc ii}] $158\,\mu\mathrm{m}$ line observations towards the Orion Nebula complex, covering M42, M43 and NGC 1973, 1975, and 1977, were obtained during 13 flights in November 2016 and February 2017 using the 14-pixel high-spectral-resolution heterodyne array of the German Receiver for Astronomy at Terahertz Frequencies (upGREAT\footnote{upGREAT is a development by the MPI f\"ur Radioastronomie and KOSMA/Universit\"at zu K\"oln, in cooperation with the DLR Institut f\"ur Optische Sensorsysteme.},\cite{Risacher2016}) onboard the Stratospheric Observatory for Infrared Astronomy (SOFIA). We obtained a fully-sampled map of a 1.2 square-degree-sized area at an angular resolution of $16\arcsec$ using the array on-the-fly (OTF) mode\footnote{The native resolution, before gridding, is $14.1\arcsec$.}. The resulting root-mean-square noise per pixel at a spectral resolution of $0.3\,\mathrm{km\,s^{-1}}$ resulted in $T_{\mathrm{mb}}\simeq 1.14\,\mathrm{K}$ after re-binning the original data with a native spectral resolution of $0.04\,\mathrm{km\,s^{-1}}$. For the baseline removal, a catalogue of splines was produced from spectra not containing an astronomical signal. and those were scaled to the astronomical data. The repeatability between flights is 7\% \citep{Higgins2021}, and we take this as an estimate of the intensity uncertainty.

\subsection{CO(2-1) observations}

We also make use of $^{12}$CO \mbox{$J$\,=\,2-1} (230.5\,GHz) and $^{13}$CO \mbox{$J$\,=\,2-1} (220.4\,GHz) line maps taken with the IRAM\,30\,m radiotelescope (Pico Veleta, Spain). The central region ($1^{\circ}\times0.8^{\circ}$) around OMC1 was originally mapped in 2008 with the HERA receiver array, and the mapping and data reduction strategies were presented by \cite{Berne2014}. These CO maps with a native angular resolution of $10.7\arcsec$ were enlarged using the new EMIR receiver and FFTS backends. The fully-sampled maps are part of the Large Program ``Dynamic and Radiative Feedback of Massive Stars''. \cite{Goicoechea2020} give details on how the old HERA and new EMIR CO maps were merged and on the conversion from antenna temperature to main-beam temperature. The typical root-mean-square noise in the \mbox{CO(2-1)} map is $T_{\mathrm{mb}}\simeq 0.16\,\mathrm{K}$ at a spectral resolution of $0.4\,\mathrm{km\,s^{-1}}$. The IRAM Large Program is an ongoing survey and we use the data as available in March 2020. Specifically, coverage in NGC 1977 is not complete and the newly observed data will be the subject of a future paper.

\subsection{Dust SEDs}

We use the dust SEDs in the Orion Nebula complex described in \cite{Pabst2020}. They use the {\it Herschel}/PACS and SPIRE photometric images of the dust FIR continuum to determine the effective dust temperature and the dust optical depth, and fit a modified blackbody,
\begin{align}
I_{\lambda} = B(\lambda,T_{\rm d})\,\left[1-\exp\left(-\tau_{160}\left(\frac{160\,\mu\mathrm{m}}{\lambda}\right)^{\beta}\right)\right], \label{eq.I}
\end{align}
with dust emissivity index $\beta=2$ to the three PACS $70\,\mu\mathrm{m}$, $100\,\mu\mathrm{m}$, and $160\,\mu\mathrm{m}$ bands, and three SPIRE $250\,\mu\mathrm{m}$, $350\,\mu\mathrm{m}$, and $500\,\mu\mathrm{m}$ bands, convolved and re-gridded to the resolution of the SPIRE $500\,\mu\mathrm{m}$ band with $36\arcsec$ angular resolution and a pixel size of $14\arcsec$. The brightest regions around the star-forming cores are saturated in the {\it Herschel} bands and hence excluded from our analysis.

The SED fits are biased towards the warm dust emission, and thus tend to underestimate the dust optical depth where colder dust is mixed in the line of sight. Moreover, the dust optical depth and dust temperature depend crucially on the exact value of $\beta$. This is examined in more detail in Paper II. Summarizing, for instance, changing $\beta$ to 1.5 in the SED fits reduces the dust optical depth by 50\%. The integrated FIR intensity is less sensitive: it varies by less than 20\%. The general behavior of the correlations identified in this study is not affected. It is worth noting in this context, however, that the flux calibration uncertainty in the PACS bands, that dominate our SED fits, is less than 7\% \citep{Balog2014}. We obtain the total FIR intensity by integrating the obtained SEDs in the range $40\text{-}500\,\mu\mathrm{m}$.

The SED fits mainly serve to determine the properties of large dust grains in radiative equilibrium. Stochastically heated UV-irradiated small dust grains might affect our dust SEDs at the shortest FIR wavelenghts. To estimate the contribution of very small grains (VSGs) to the PACS $70\,\mu\mathrm{m}$ band we compute the intensity ratio of this band with the {\it Spitzer}/Multi-band Imaging Photometer for {\it Spitzer} (MIPS) $24\,\mu\mathrm{m}$ band. At the moderate temperatures over most of the region, a high $24\,\mu\mathrm{m}$/$70\,\mu\mathrm{m}$ ratio indicates a large contribution by VSGs. Dust inside H\,{\sc ii} regions is heated by trapped Ly$\alpha$ radiation \citep{Salgado2016} and thus has a high $24\,\mu\mathrm{m}$/$70\,\mu\mathrm{m}$ ratio. We conclude that the contribution of VSGs is small in the line of sight towards the extended shell of the Orion Nebula. Only in the Veil bubble interior, that is filled with the hot plasma generated by the stellar wind from $\theta^1$ Ori C \citep{Guedel2008}, contribution by VSGs might be significant. We surmise that emission from this region is dominated by dust collisionally heated by hot electrons and cooled through IR emission \citep{Dwek1992}. We stress that, examining the optical depth and temperature maps, the FIR emission in these sight lines is dominated by the colder material in the molecular cloud in the background.

\subsection{Ancillary photometric data}

We make use of H$\alpha$ observations obtained by the Wide Field Imager (WFI) on the European Southern Observatory (ESO) telescope at La Silla \citep{DaRio2009} and the ESO/Digitized Sky Survey 2 (DSS-2) red-band image. We use the H$\alpha$ image obtained with the Very Large Telescope (VLT)/Multi Unit Spectroscopic Explorer (MUSE) towards the Huygens Region \citep{Weilbacher2015} to calibrate the WFI and DSS-2 images. The DSS-2 image is saturated in the inner EON, but that region is well covered by the WFI image.

Moreover, we employ the {\it Spitzer}/Infrared Array Camera (IRAC) $8\,\mu\mathrm{m}$ image of the Orion Nebula Complex \citep[described in][]{Megeath2012} to trace emission from poly\-cyclic aromatic hydrocarbons (PAHs). The $8\,\mu\mathrm{m}$ band contains predominantly emission from ionized PAH species. We apply the surface brightness correction factor, described in the IRAC handbook, of 0.74 to the archival image. This induces a calibration uncertainty of about 10\%, according to the IRAC handbook.

\begin{figure}[ht]
\includegraphics[width=0.5\textwidth, height=0.5\textwidth]{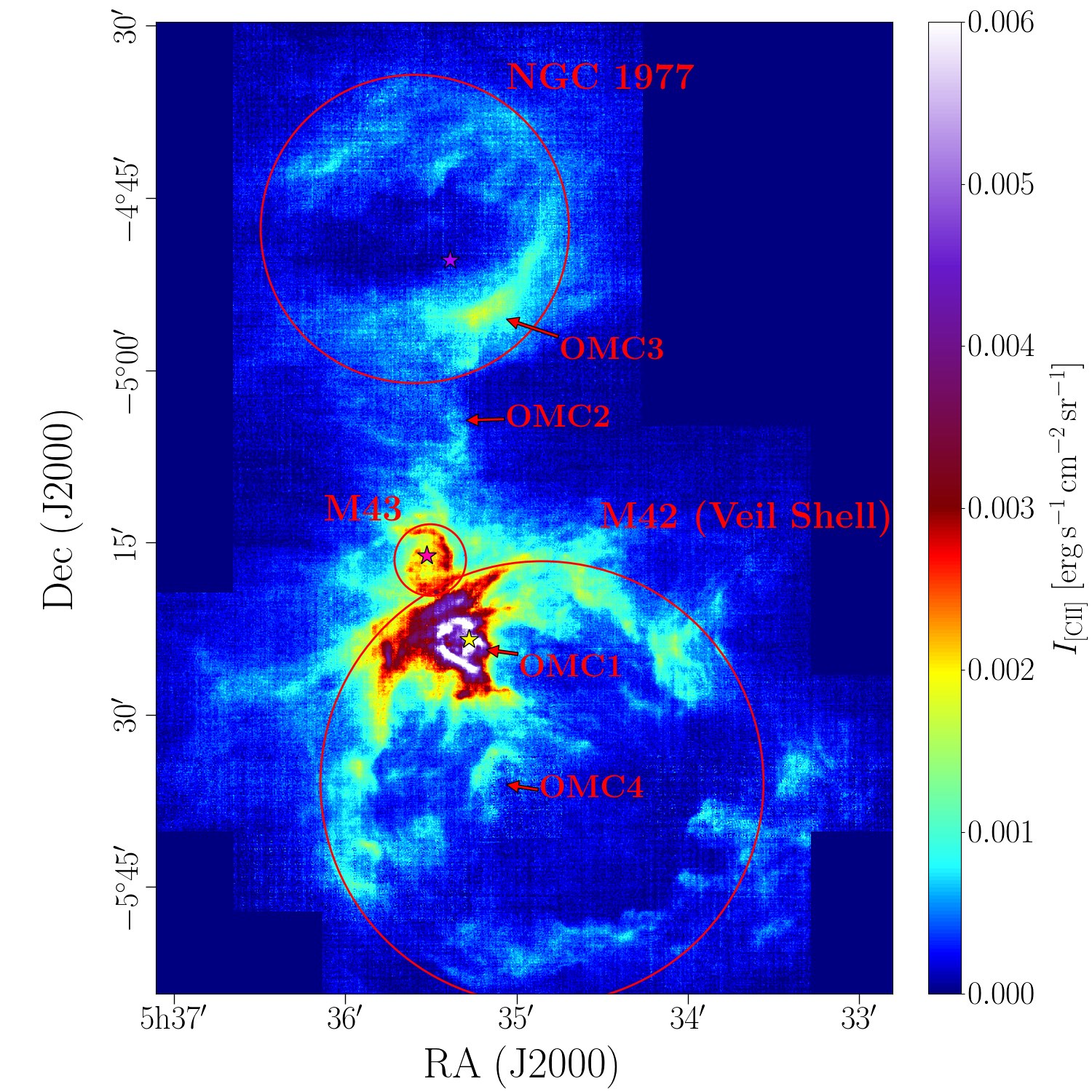}
\caption{[C\,{\sc ii}] line-integrated intensity ($v_{\mathrm{LSR}}=-10\,\text{-}\,+20\,\mathrm{km\,s^{-1}}$) from Orion A at its original resolution ($16\arcsec$). The red circles delineate the three distinct shells of M42, M43, and NGC 1977. The stars indicate the most massive stars within each region: $\theta^1$ Ori C (yellow) in M42, NU Ori (pink) in M43, and 42 Orionis (purple) in NGC 1977. Arrows indicate the positions of the four molecular cores OMC1-4 along the Integral-Shaped Filament.}
\label{Fig.map}
\end{figure}

\begin{figure*}[ht]
\includegraphics[width=\textwidth, height=0.67\textwidth]{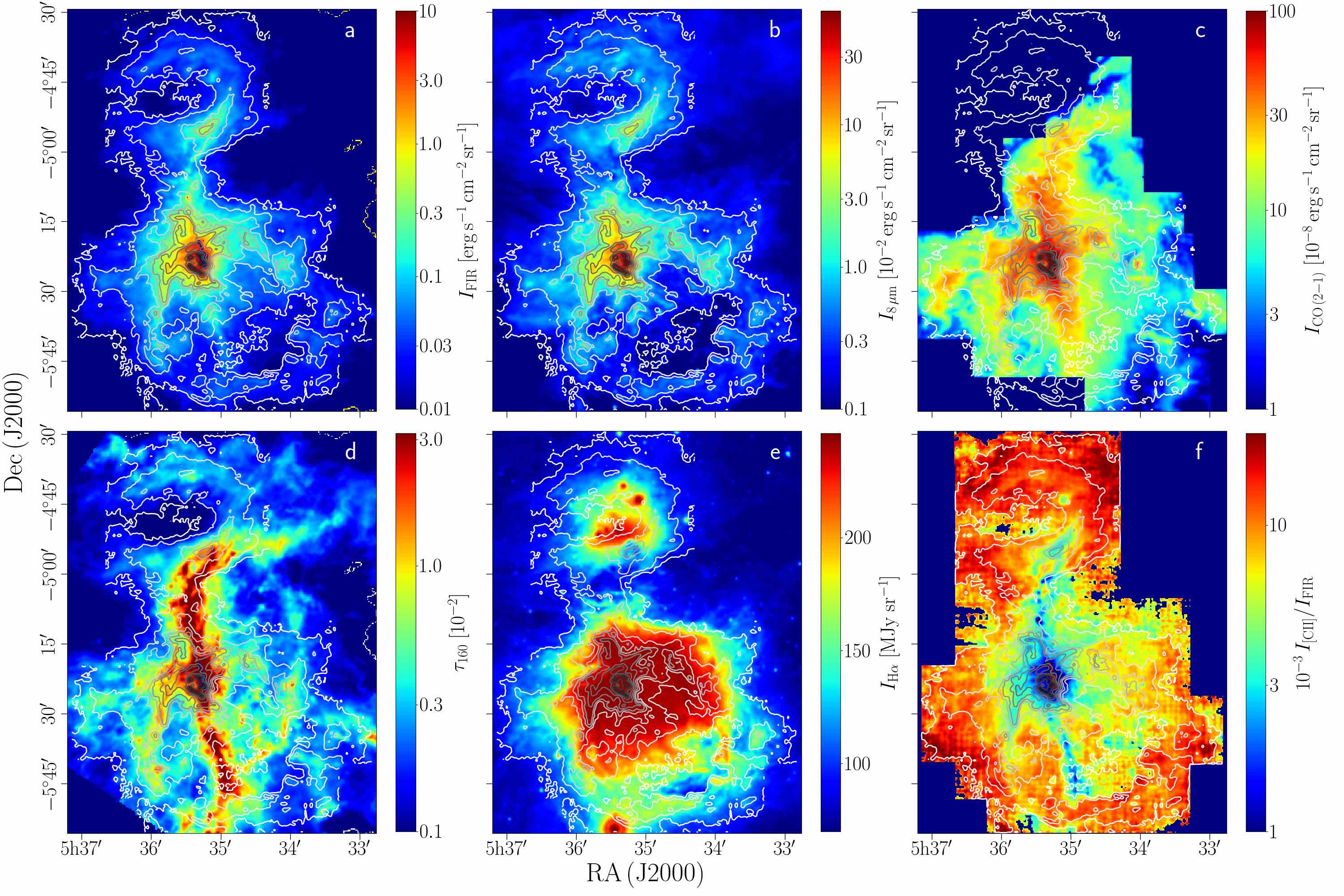}
\caption{Various gas and dust tracers with [C\,{\sc ii}] line-integrated intensity ($v_{\mathrm{LSR}}=-10\text{-}20\,\mathrm{km\,s^{-1}}$) in contours (from white to black: $1.5, 3.5, 7.0, 10.5, 14.0, 20.0, 30.0,  40.0, 50.0, 60.0\times 10^{-4}\,\mathrm{erg\,s^{-1}\,cm^{-2}\,sr^{-1}}$): {\it a)} FIR intensity, {\it b)} IRAC $8\,\mu\mathrm{m}$ intensity, {\it c)} IRAM 30m CO(2-1) intensity, {\it d)} dust optical depth $\tau_{160}$, {\it e)} DSS H$\alpha$ intensity, {\it f)} [C\,{\sc ii}]/FIR ratio. All images are convolved to the resolution of the FIR intensity image ($36\arcsec$). Note that the DSS H$\alpha$ image (panel e) is saturated at $I_{\mathrm{H\alpha}} \simeq 250\,\mathrm{MJy\,sr^{-1}}$.}
\label{Fig.contours}
\end{figure*}

\section{Analysis}

For the analysis, we convolve and re-grid all images to the resolution of the FIR intensity image, that is the poorest resolution with a beam size of $36\arcsec$ and a pixel size of $14\arcsec$.

\subsection{Global morphology of the emission}

Figure \ref{Fig.map} shows the [C\,{\sc ii}] line-integrated intensity from Orion A. The [C\,{\sc ii}]-mapped area in Orion A comprises three distinct regions. The most massive stars in the Orion Nebula complex, the Trapezium stars, are found close to the surface of OMC1. Here, they create a heavily irradiated PDR at the surface of the molecular cloud, that radiates bright [C\,{\sc ii}] emission. Also the Veil Shell, the expanding shell that is created by the stellar wind of the O7V star $\theta^1$ Ori C (the most massive of the Trapezium stars), radiates (limb-brightened) [C\,{\sc ii}] emission. In M43, we observe a limb-brightened [C\,{\sc ii}]-emitting shell structure, as well, in addition to the strongly irradiated PDR on the background molecular cloud. The shell surrounding NGC 1973, 1975, and 1977 also emits substantially in the [C\,{\sc ii}] line, the brightest part, however, being the PDR at the surface of OMC3.

In this paper, we will first focus on the global correlations of the [C\,{\sc ii}] emission with other star-formation tracers. In Paper II we will study the correlations in more detail by dividing them into several subregions. Figure \ref{Fig.contours} shows the distribution of various gas and dust tracers (FIR, $8\,\mu\mathrm{m}$, CO(2-1), $\tau_{160}$, H$\alpha$ intensities, [C\,{\sc ii}]/FIR ratio) in the Orion Nebula complex with the [C\,{\sc ii}] line-integrated intensity in contours. While the morphology of the FIR, $8\,\mu\mathrm{m}$ and [C\,{\sc ii}] emission is very similar, the CO(2-1) intensity and the dust optical depth $\tau_{160}$ exhibit a different morphology, tracing the dense gas in the star-forming filament (the Integral-Shaped Filament, ISF). H$\alpha$ emission traces the ionized gas and is concentrated inside the limb-brightened edges that light up in FIR, $8\,\mu\mathrm{m}$ and [C\,{\sc ii}] emission. The [C\,{\sc ii}]/FIR ratio traces the gas heating efficiency and generally increases away from the luminous [C\,{\sc ii}] sources close to the central stars. Dense filamentary structures, like the ISF, also possess low [C\,{\sc ii}]/FIR ratios, as the FIR emission is dominated by large columns of cool dust.

\begin{figure*}[tb]
\begin{minipage}{0.49\textwidth}
\includegraphics[width=\textwidth, height=0.67\textwidth]{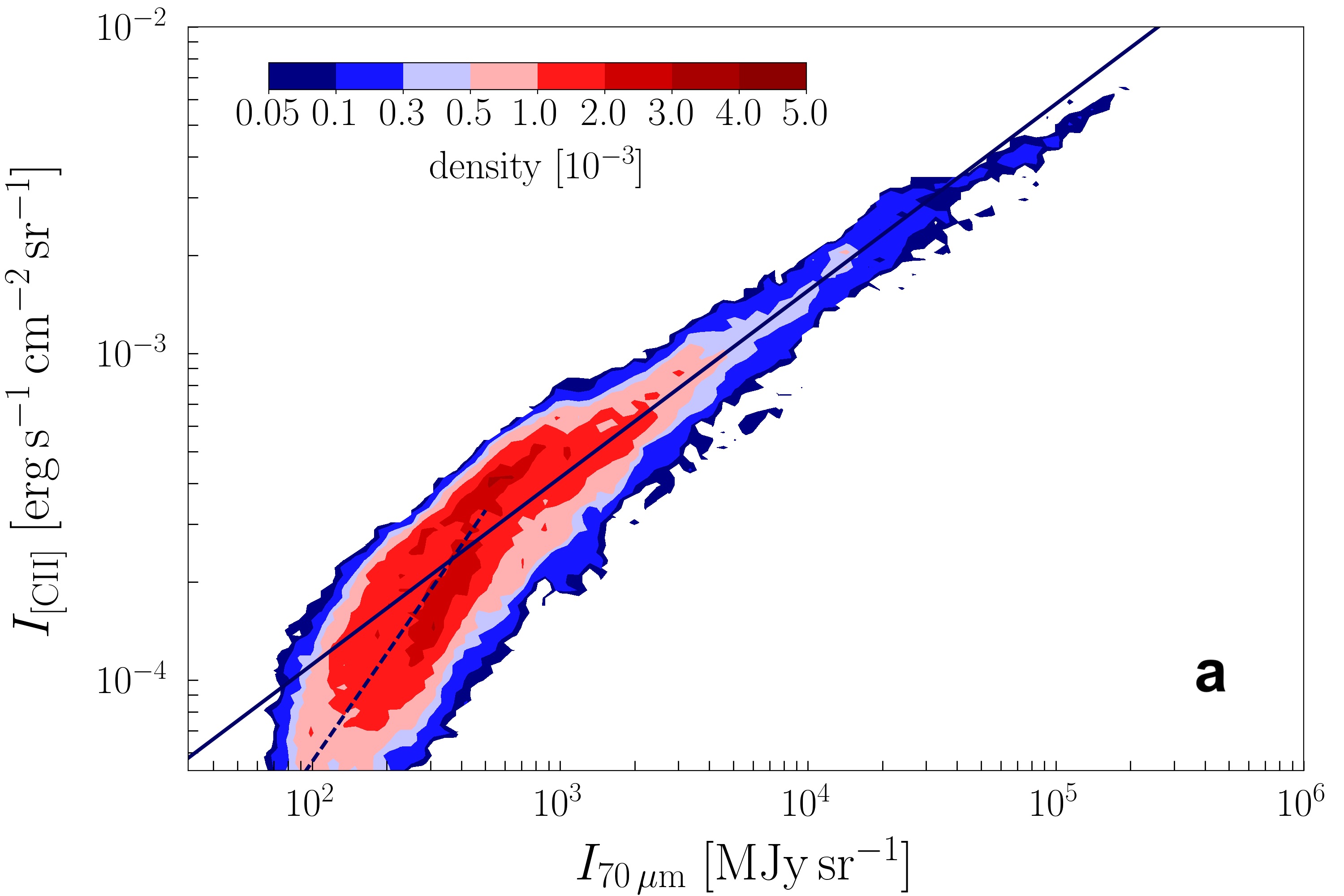}
\end{minipage}
\begin{minipage}{0.49\textwidth}
\includegraphics[width=\textwidth, height=0.67\textwidth]{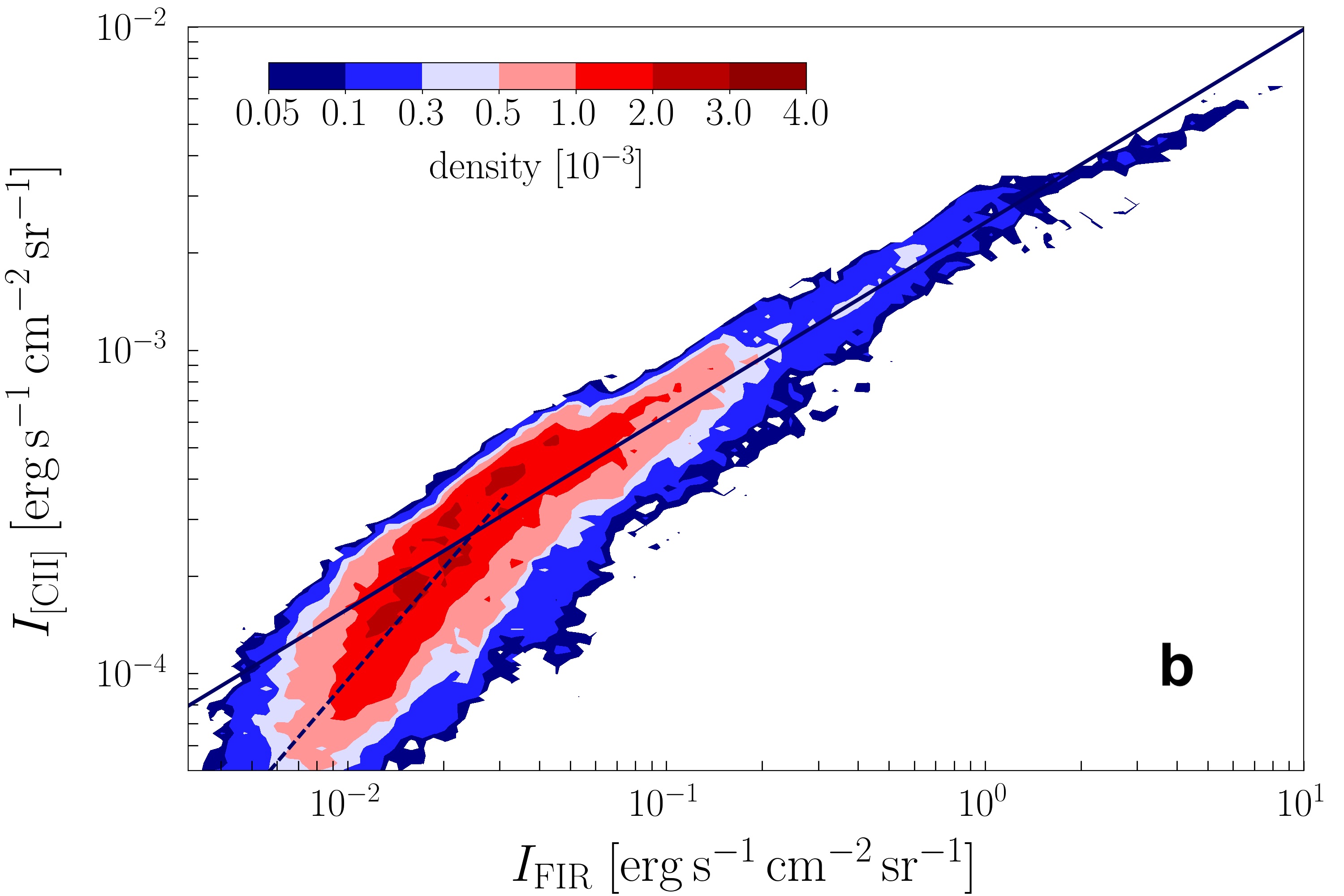}
\end{minipage}

\begin{minipage}{0.49\textwidth}
\includegraphics[width=\textwidth, height=0.67\textwidth]{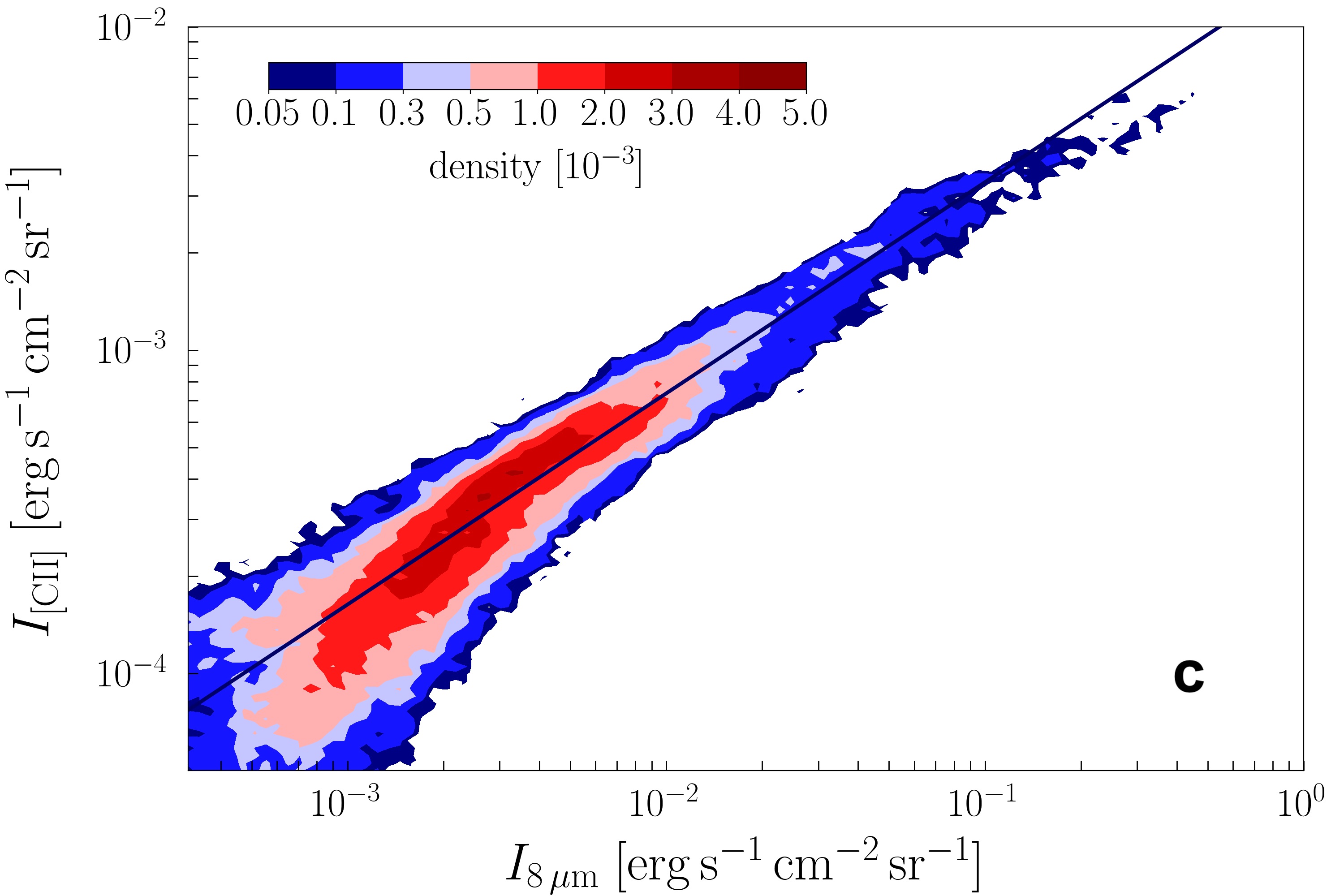}
\end{minipage}
\begin{minipage}{0.49\textwidth}
\includegraphics[width=\textwidth, height=0.67\textwidth]{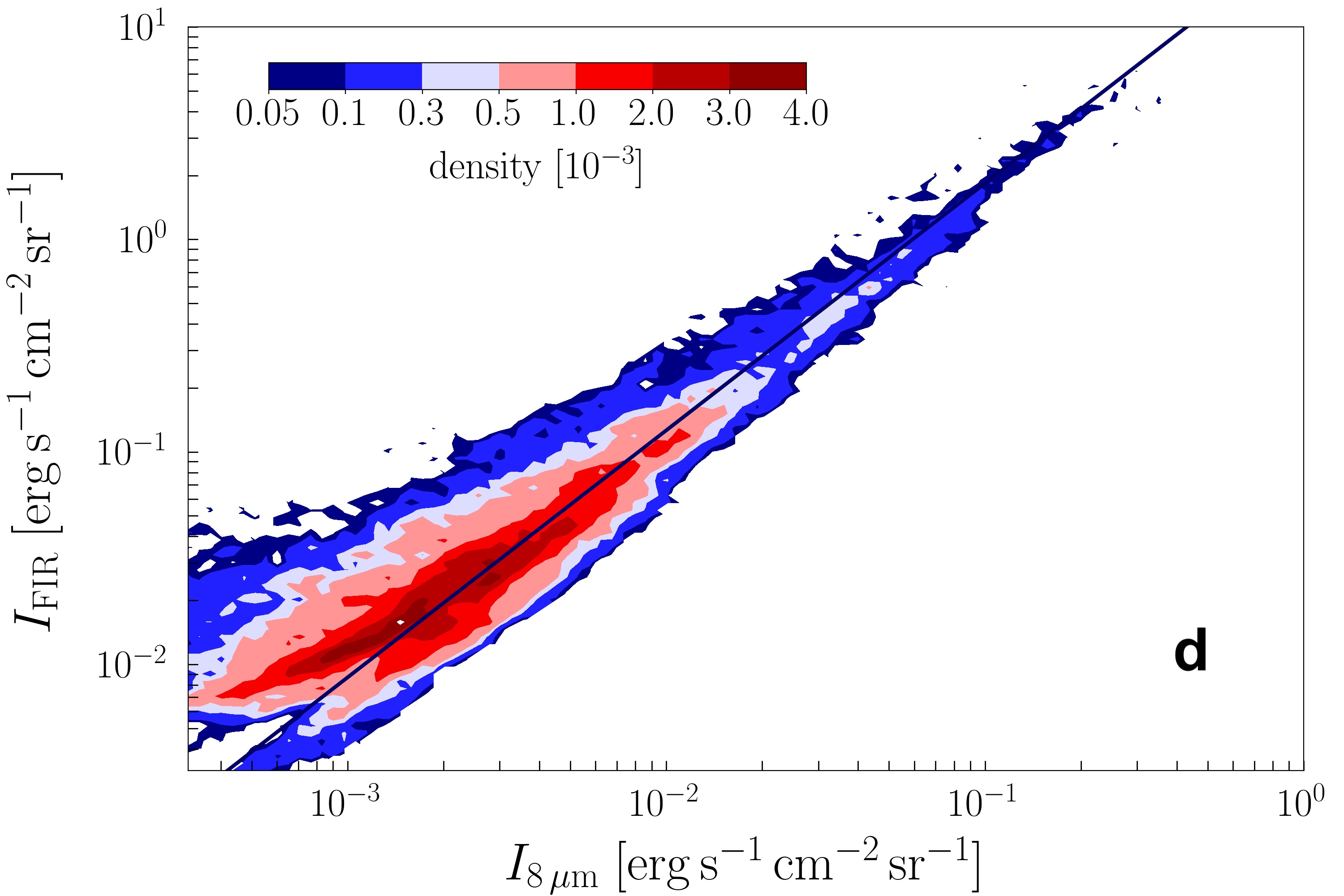}
\end{minipage}

\begin{minipage}{0.49\textwidth}
\includegraphics[width=\textwidth, height=0.67\textwidth]{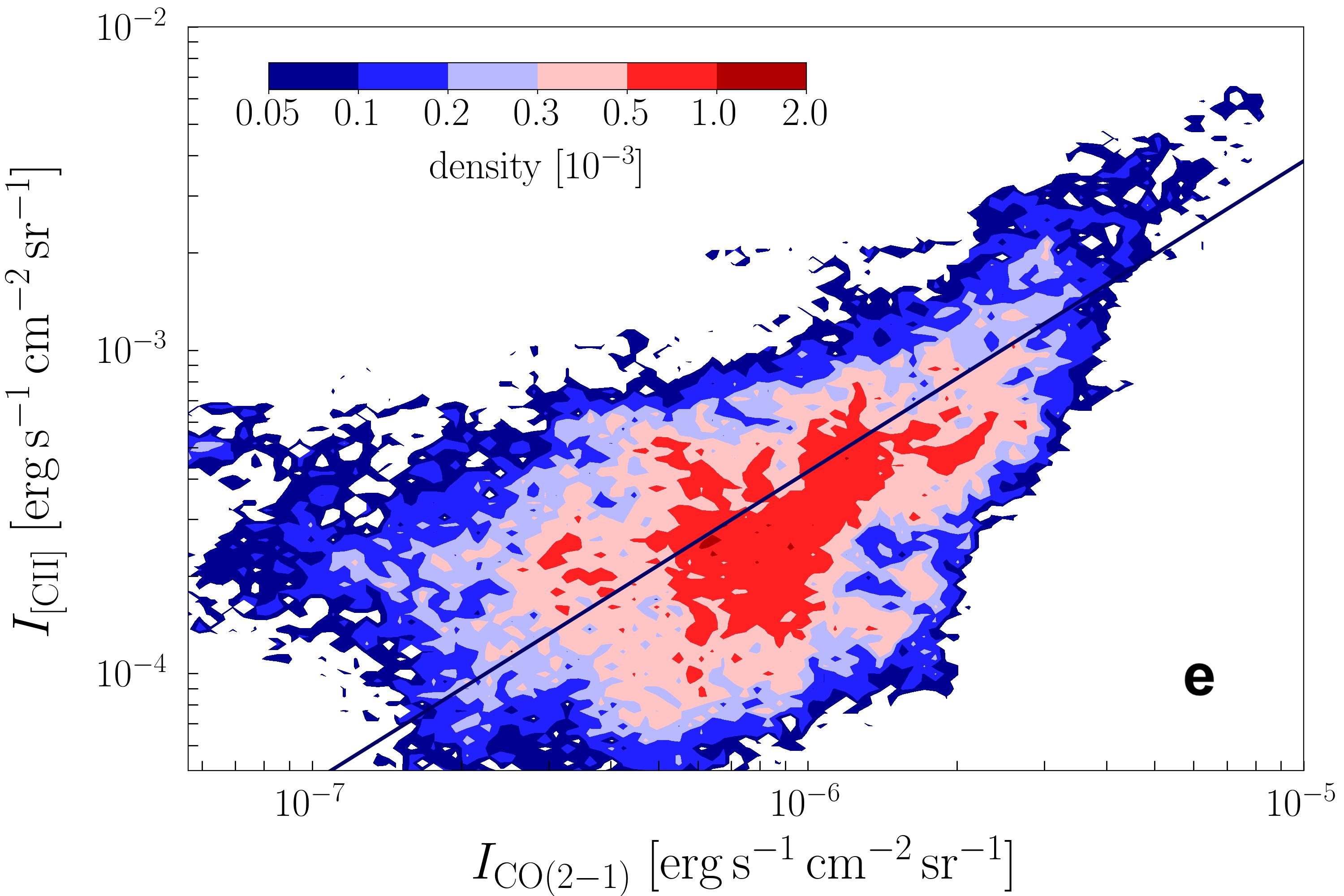}
\end{minipage}
\begin{minipage}{0.49\textwidth}
\includegraphics[width=\textwidth, height=0.67\textwidth]{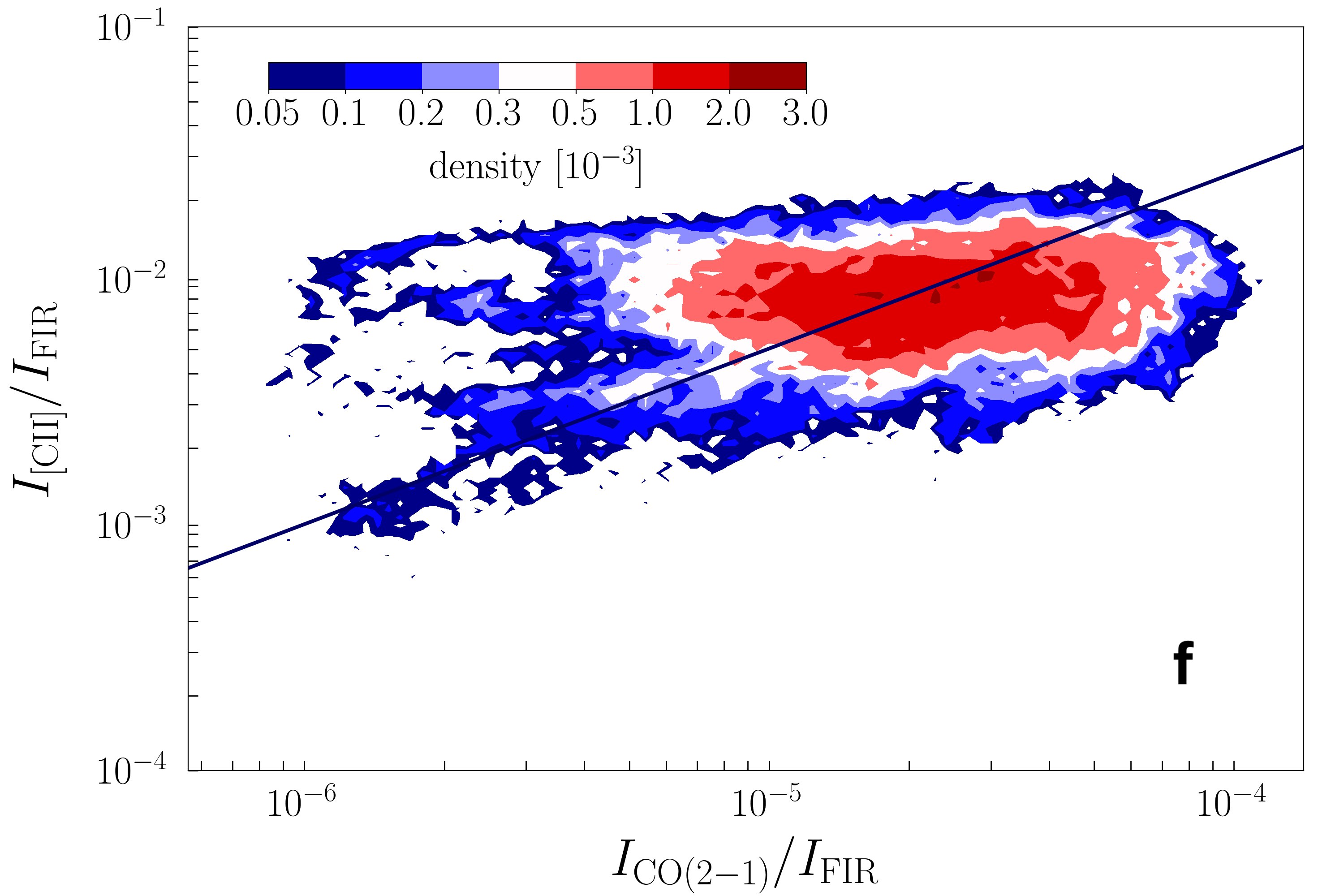}
\end{minipage}
\caption{Correlations of all map points as point-density plots: {\it a)} [C\,{\sc ii}]-$70\,\mu\mathrm{m}$, {\it b)} [C\,{\sc ii}]-FIR, {\it c)} [C\,{\sc ii}]-$8\,\mu\mathrm{m}$, {\it d)} FIR-$8\,\mu\mathrm{m}$, {\it e)} [C\,{\sc ii}]-CO(2-1), {\it f)} [C\,{\sc ii}]/FIR-CO(2-1)/FIR. The regression curve with parameters of Table \ref{Tab.fits_global} is drawn in the respective panels (solid lines; dashed lines in panels a and b are the regression curves in the low-intensity regime).}
\label{Fig.densities}
\end{figure*}

\subsection{Correlation plots of gas and dust tracers}
\label{Sec.correlations}

\begin{table}[ht]
\addtolength{\tabcolsep}{-3pt}
\def\arraystretch{1.2}
\caption{Summary of the point-density correlation plots in Fig. \ref{Fig.densities}.}
\begin{tabular}{ll|cccc}
\hline\hline
x & y & $a$ & $b$ & $\rho$ & $rms\,\mathrm{[dex]}$\\ \hline
$70\,\mu\mathrm{m}$ & {[C\,{\sc ii}]} & $0.57\pm 0.31$\tablefootmark{a} & $-5.10\pm 1.02$ & 0.90 & 0.13\\
FIR & {[C\,{\sc ii}]} & $0.60\pm 0.28$\tablefootmark{b} & $-2.61\pm 0.29$ & 0.90 & 0.12\\
$8\,\mu\mathrm{m}$ & {[C\,{\sc ii}]} & $0.65\pm 0.30$ & $-1.83\pm 0.64$ & 0.93 & 0.10\\
$8\,\mu\mathrm{m}$ & FIR & $1.16\pm 0.37$ & $1.43\pm 0.77$ & 0.94 & 0.16\\
CO(2-1) & {[C\,{\sc ii}]} & $0.96\pm 0.63$ & $2.39\pm 3.89$ & 0.53 & 0.37\\
$\frac{\mbox{CO(2-1)}}{\mbox{FIR}}$ & $\frac{\mbox{[C\,{\sc ii}]}}{\mbox{FIR}}$ & $0.71\pm 0.60$ & $1.24\pm 2.85$ & 0.34 & 0.29\\
\end{tabular}
\tablefoot{The regression fitted to the data in a weighted least-square fit is $\log_{10} y = a \log_{10} x + b$. Fit in $I_{\mathrm{[C\,\textsc{ii}]}}>3\sigma\simeq 5\times 10^{-5}\,\mathrm{erg\,s^{-1}\,cm^{-2}\,sr^{-1}}$, $I_{70\,\mu\mathrm{m}}>5\times 10^{2}\,\mathrm{MJy\,sr^{-1}}$, $I_{8\,\mu\mathrm{m}}>3\times 10^{-3}\,\mathrm{erg\,s^{-1}\,cm^{-2}\,sr^{-1}}$, and/or $I_{\mathrm{FIR}}>3\times 10^{-2}\,\mathrm{erg\,s^{-1}\,cm^{-2}\,sr^{-1}}$. $\rho$ is the Pearson correlation coefficient, $rms$ is the root-mean-square of the residual of the fit.
\tablefoottext{a}{For $I_{70\,\mu\mathrm{m}}<5\times 10^{2}\,\mathrm{MJy\,sr^{-1}}$ we obtain $a\simeq 1.10\pm 0.68$ and $b\simeq -6.45\pm 1.61$ with $\rho\simeq 0.67$ and $rms\simeq 0.18$.}
\tablefoottext{b}{For $I_{\mathrm{FIR}}<3\times 10^{-2}\,\mathrm{erg\,s^{-1}\,cm^{-2}\,sr^{-1}}$ we obtain $a\simeq 1.15\pm 0.76$ and $b\simeq -1.72\pm 1.37$ with $\rho\simeq 0.71$ and $rms\simeq 0.17$.}}
\label{Tab.fits_global}
\end{table}

Figure \ref{Fig.densities} shows the point-by-point correlation between the different gas and dust tracers as density plots. Table \ref{Tab.fits_global} summarizes the power-law fits we find for each correlation, using the ordinary least-squares (OLS) bisector method \citep{Isobe1990} on the logarithmic scale. We have included all points that lie above the $3\sigma\simeq 8\,\mathrm{K\;km\,s^{-1}}\simeq 5\times 10^{-5}\,\mathrm{erg\,s^{-1}\,cm^{-2}\,sr^{-1}}$ threshold of the [C\,{\sc ii}] line-integrated intensity (after convolution). We do not apply this threshold to the FIR-$8\,\mu\mathrm{m}$ correlation, but we restrict both maps to the coverage of the [C\,{\sc ii}] map. We subtract an offset of $2\times 10^{-3}\,\mathrm{erg\,s^{-1}\,cm^{-2}\,sr^{-1}}$ from the IRAC $8\,\mu\mathrm{m}$ data. However, at low intensities we find 70\,$\mu$m and FIR emission that does not correspond to [C\,{\sc ii}] emission. We fit this regime ($I_{70\,\mu\mathrm{m}}<5\times 10^{2}\,\mathrm{MJy\,sr^{-1}}$ and $I_{\mathrm{FIR}}<3\times 10^{-2}\,\mathrm{erg\,s^{-1}\,cm^{-2}\,sr^{-1}}$, respectively) separately. We point out that this low-intensity regime is problematic because of calibration uncertainties in the PACS bands, hence we will refrain from a detailed analysis thereof. The same holds for the IRAC $8\,\mu\mathrm{m}$ map, where we cannot exclude that the offset we subtract has a physical reason. We caution that this has to be taken into account when estimating the [C\,{\sc ii}] intensity from the IRAC $8\,\mu\mathrm{m}$ intensity at low surface brightness.

The [C\,{\sc ii}] intensity is tightly correlated with the intensity in the PACS $70\,\mu\mathrm{m}$ band (Fig. \ref{Fig.densities}a), the FIR intensity (Fig. \ref{Fig.densities}b), and with the intensity in the IRAC $8\,\mu\mathrm{m}$ band (Fig. \ref{Fig.densities}c). The FIR intensity is also tightly correlated with the intensity in the IRAC $8\,\mu\mathrm{m}$ band (Fig. \ref{Fig.densities}d). The Pearson correlation coefficients ($\rho$) in panels a to d are about 0.9, while the scatter around the correlations is of the order $0.1\text{-}0.2\,\mathrm{dex}$. Most important to note is that the correlation of the [C\,{\sc ii}] intensity with each of the $70\,\mu\mathrm{m}$, FIR (in the high-intensity regime), and $8\,\mu\mathrm{m}$ intensity is not linear, but scales with those intensities to a power of less than unity. The FIR-$8\,\mu\mathrm{m}$ correlation deviates only slightly from linearity. At low FIR intensity, the [C\,{\sc ii}] intensity and the $8\,\mu\mathrm{m}$ intensity are largely independent of the FIR intensity. 

The [C\,{\sc ii}]-CO(2-1) correlations are more complex and require a more thorough analysis, which we will return to and follow up in Paper II. The [C\,{\sc ii}]-low-J CO relationship, either as absolute intensities or normalized to the FIR intensity, is frequently used to infer physical conditions in Galactic and extragalactic star-forming regions \citep{Wolfire1989, Stacey1991, Herrmann1997}, which is why we include it here. Collectively, the [C\,{\sc ii}]-CO(2-1) correlations in Orion A (Figs. \ref{Fig.densities}e and f) do not follow a simple power-law trend.

\begin{figure}[tb]
\begin{minipage}{0.49\textwidth}
\includegraphics[width=\textwidth, height=0.67\textwidth]{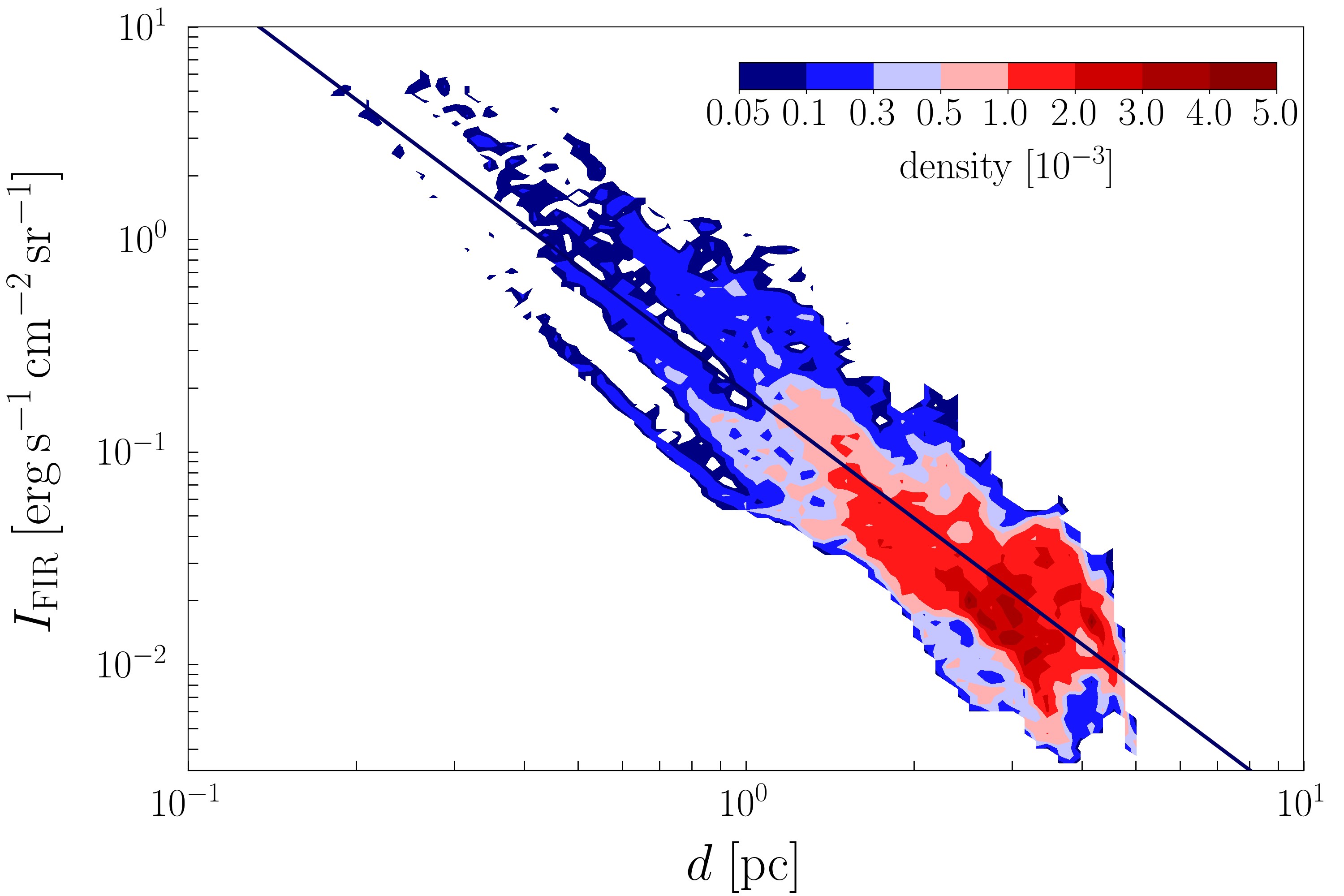}
\end{minipage}
\caption{FIR intensity in M42 versus (projected) distance from $\theta^1$ Ori C. The solid line is a power-law fit with $I_{\mathrm{FIR}}=b d^a$ and $a\simeq -1.97\pm 0.55$, $b\simeq 10^{-0.72\pm 0.24}\,\mathrm{erg\,s^{-1}\,cm^{-2}\,sr^{-1}}$.}
\label{Fig.FIR-d}
\end{figure}

To estimate the FUV radiation field, we discuss briefly the FIR intensity as function of the distance from the massive stars\footnote{We assume a distance of $414\,\mathrm{pc}$ to the Orion Nebula \citep{Menten2007} to be consistent with \cite{Pabst2019,Pabst2020}. More recent observations suggest a slightly lower value \citep{Kounkel2017,Grossschedl2018}.}. Figure \ref{Fig.FIR-d} shows the dependence of the FIR intensity on distance from the central star $\theta^1$ Ori C in M42. We find a decline of the FIR intensity in accordance with geometric dilution, that is decreasing with the distance squared. The FIR intensity scales directly with the incident FUV radiation field, given as $G_0$ in units of the Habing field, typically $G_0\simeq I_{\mathrm{FIR}}/2/1.3\times 10^{-4}\,\mathrm{erg\,s^{-1}\,cm^{-2}\,sr^{-1}}$, where the factor $1/2$ accounts for the absorption of visible photons by dust \citep{HollenbachTielens1999}. However, geometry and line-of-sight effects are important, most of the sight lines being towards limb-brightened cloud edges. The true FUV field may be found by the lower intensities in the correlation of Fig. \ref{Fig.FIR-d}, as those correspond to the face-on background cloud and are not affected by limb brightening. This yields $G_0\simeq 500$ at a distance of $1\,\mathrm{pc}$. Also following the prescription of \cite{Meixner1992}, $G_0\simeq L_{\mathrm{FIR}}/(4\pi S)/2/1.3\times 10^{-4}\,\mathrm{erg\,s^{-1}\,cm^{-2}\,sr^{-1}}$, where $S$ is the surface that is exposed to the FUV radiation field, we obtain $G_0\simeq 500$ at a distance of 1\,pc from the FIR luminosity of the edge-on eastern shell\footnote{We sum over an area of $3\arcmin\times 3\arcmin$ in the limb-brightened eastern edge, obtaining $L_{\mathrm{FIR}}\simeq 1.8\times 10^3\,L_{\sun}$. At a distance of 414\,pc, $3\arcmin\simeq 0.36\,\mathrm{pc}$. The column exposed to the FUV radiation field has an approximate length of $r/2\simeq 1.3\,\mathrm{pc}$, where $r\simeq 2.5\,\mathrm{pc}$ is the radius of the shell structure. Hence, $S\simeq 0.36\,\mathrm{pc}\times 1.3\,\mathrm{pc}$.}. From the total FIR luminosity of M42, $1.5\times 10^5\,L_{\sun}$ excluding the BN/KL and Orion S region, we estimate $G_0\simeq 1500$ at a distance of 1\,pc. While, in principle, the incident radiation field can be estimated from the stellar luminosity and the (true) distance from the illuminating source, oftentimes only the projected distance is known. In the case of the Orion Nebula, the detailed geometry of the irradiated gas and dust is complex and also the incident angle of the FUV radiation field with respect to the surface normal has to be taken into account \citep[cf. Fig. 13 in][]{ODell2010}. Moreover, the FUV radiation field in the Veil Shell may be attenuated by dust in the H\,{\sc ii} region and possibly in gas layers between the Trapezium stars and the Veil Shell \citep[cf.][]{VanderWerf2013, Abel2019}.

\section{Discussion}

The Orion A molecular cloud towards the Orion Nebula (NGC 1976 or M42) is the closest site of ongoing massive-star formation, and thus a perfect template to resolve the spatial distribution of the [C\,{\sc ii}] emission. At a distance of $414\pm 7\mathrm{pc}$ from the earth \citep{Menten2007}, it covers about one square degree on the sky. To the north of the Orion Nebula lies De Mairan's Nebula (NGC 1982 or M43) and the Running-Man Nebula with NGC 1973, 1975, and 1977. The Orion Nebula itself is illuminated by the massive stars in the Trapezium cluster, the dominant star being the O7V star $\theta^1$ Ori C at only  $\sim 0.3\,\mathrm{pc}$ from the molecular cloud. According to the blister model \citep{Zuckerman1973,Balick1974}, the environing H\,{\sc ii} region, the so-called Huygens Region, is located in front of the background Orion A molecular cloud. This very central part of the cloud includes the Orion Molecular Core 1 (OMC1) that hosts the embedded massive-star forming clumps \mbox{Orion\,BN/KL} and \mbox{Orion\,S} \citep[e.g.,][]{Genzel1989}. OMC1 also hosts the famous Orion Bar, a FUV-illuminated edge of the molecular cloud, and an archetypical example of a strongly irradiated PDR ($G_0\simeq 10^4$) that has been subject of many detailed studies \citep[e.g.,][]{Tielens1993,Goicoechea2016}. The H\,{\sc ii} regions of M43 and NGC 1977 are ionized by the B0.5V star NU Ori and the B1V star 42 Ori, respectively. All these regions are subject to ongoing dynamical evolution. M42 is surrounded by a stellar-wind driven shell expanding at  $13\,\mathrm{km\,s^{-1}}$ \citep{Pabst2019}, filled with a photo-ionized medium \citep{ODell2010} and an X-ray emitting hot plasma \citep{Guedel2008}. Both M43 and NGC 1973, 1975, and 1977 are surrounded by shells that expand under the influence of the contained over-pressurized ionized gas \citep{Pabst2020}.

\subsection{Comparison with Orion B}
\label{sec.OrionB}

An early spatial correlation study of the central OMC1/Huygens region ($\sim 85\,\mathrm{arcmin}^2$) was carried out by \citet{Goicoechea2015} using Herschel/HIFI. This region is very bright in the FIR and shows very low [C\,{\sc ii}]/FIR intensity ratios (\mbox{approaching 10$^{-4}$}) reminiscent of the \mbox{``[C\,{\sc ii}] deficit''} seen in local ULIRGs. These authors concluded that the low luminosity ratios in this particular region are produced by the very large column density of warm dust throughout the star-forming core. Here, we expand this correlation study to much larger spatial scales, encompassing a much wider range of physical conditions, dust column densities, and FUV radiation illumination. In order to include regions of very low [C\,{\sc ii}] surface brightness into the analysis, and to study the role of different environment/stellar-illumination conditions, we also incorporate the [C\,{\sc ii}] emission mapped by SOFIA around the Horsehead Nebula and adjacent IC 434 H\,{\sc ii} region \citep[located in Orion B, about $3^{\circ}$ north of Orion A;][]{Pabst2017,Bally2018}. This region in Orion B is illuminated by the multiple star system $\sigma$ Ori, dominated by a massive star of spectral type O9.5V and located $\sim 3\,\mathrm{pc}$ (in projection) from the Horsehead, producing $G_0\simeq 100$ at the surface of the molecular cloud \citep[e.g.,][]{Abergel2003}.

\begin{figure*}[tb]
\begin{minipage}{0.49\textwidth}
\includegraphics[width=\textwidth, height=0.67\textwidth]{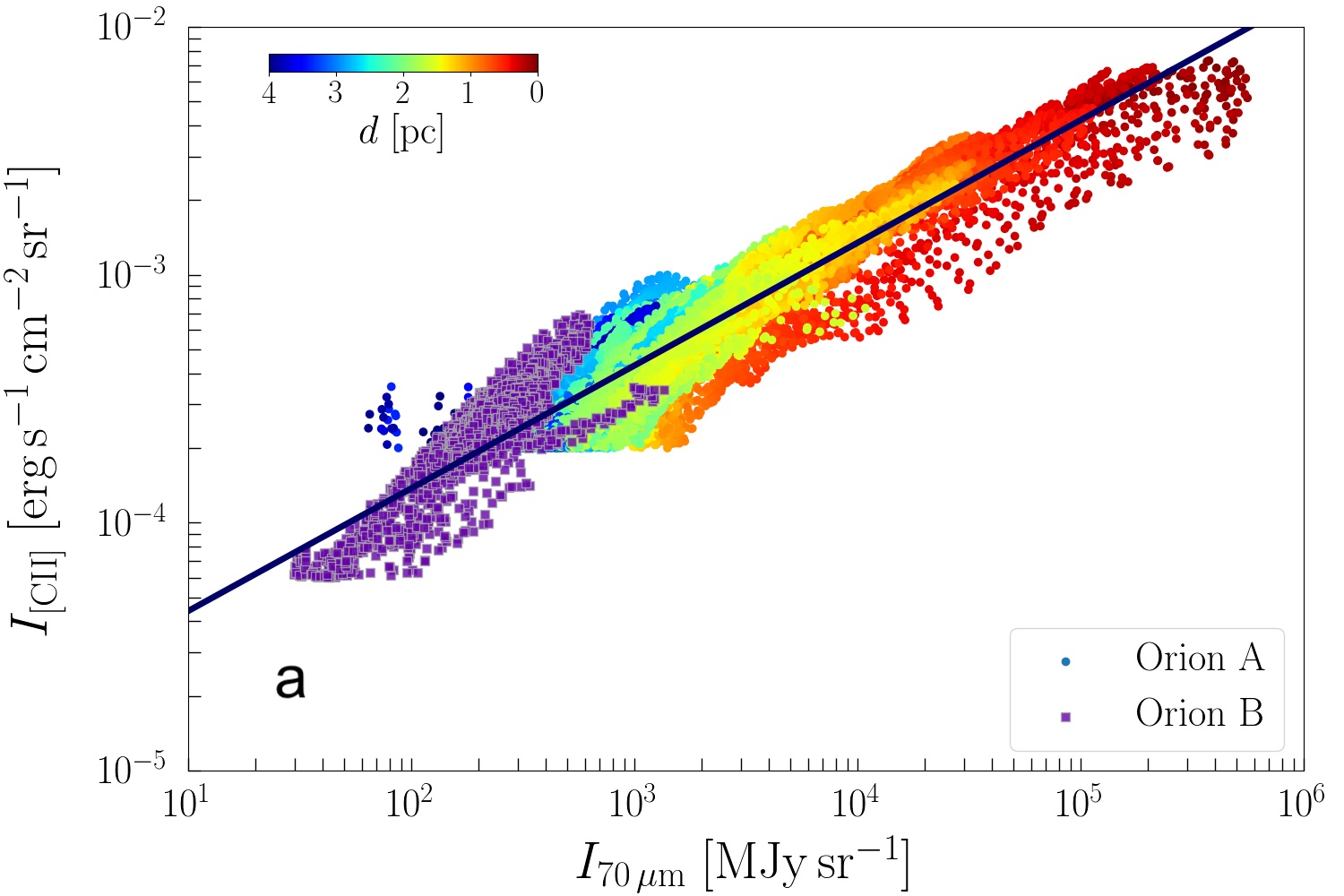}
\end{minipage}
\begin{minipage}{0.49\textwidth}
\includegraphics[width=\textwidth, height=0.67\textwidth]{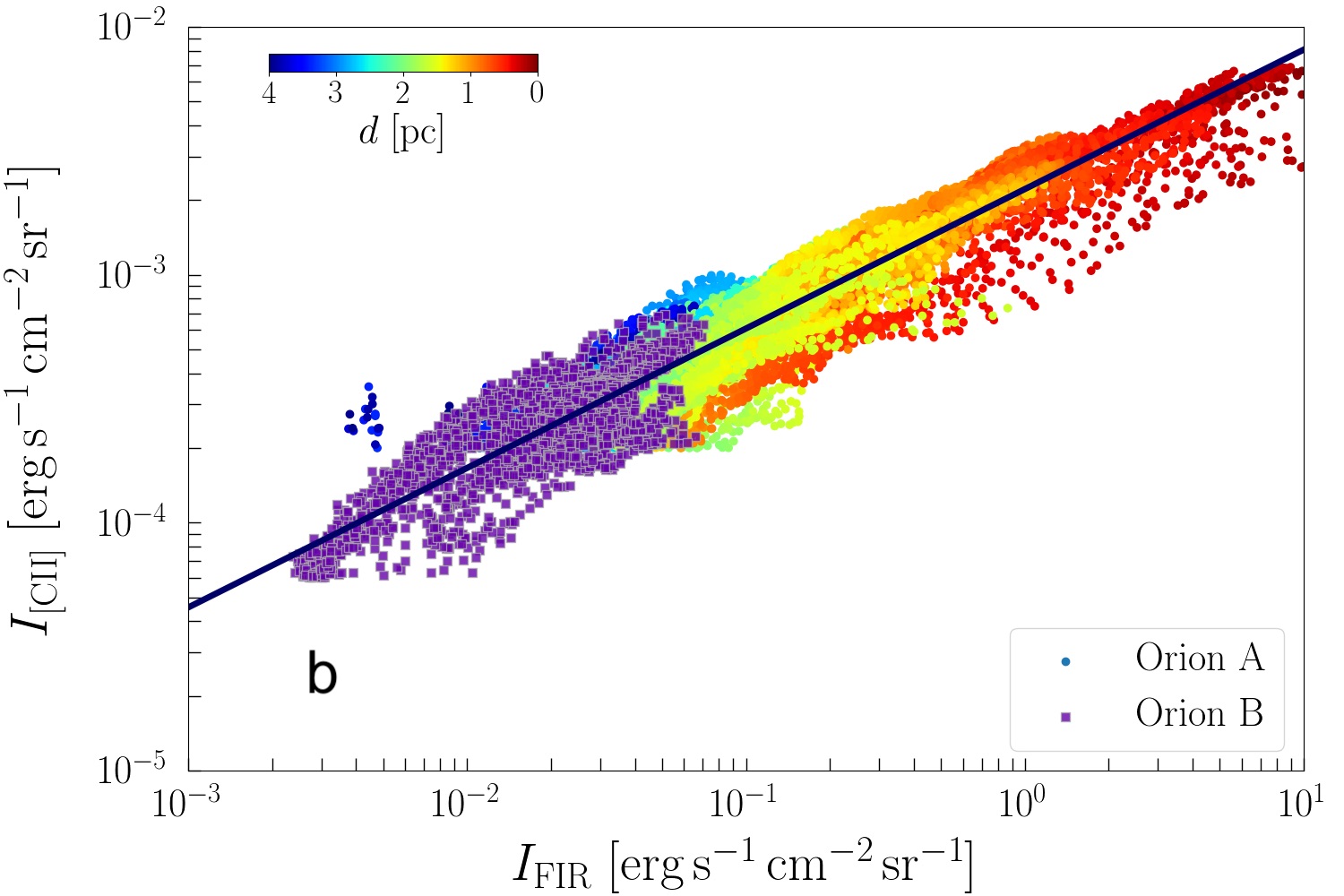}
\end{minipage}

\begin{minipage}{0.49\textwidth}
\includegraphics[width=\textwidth, height=0.67\textwidth]{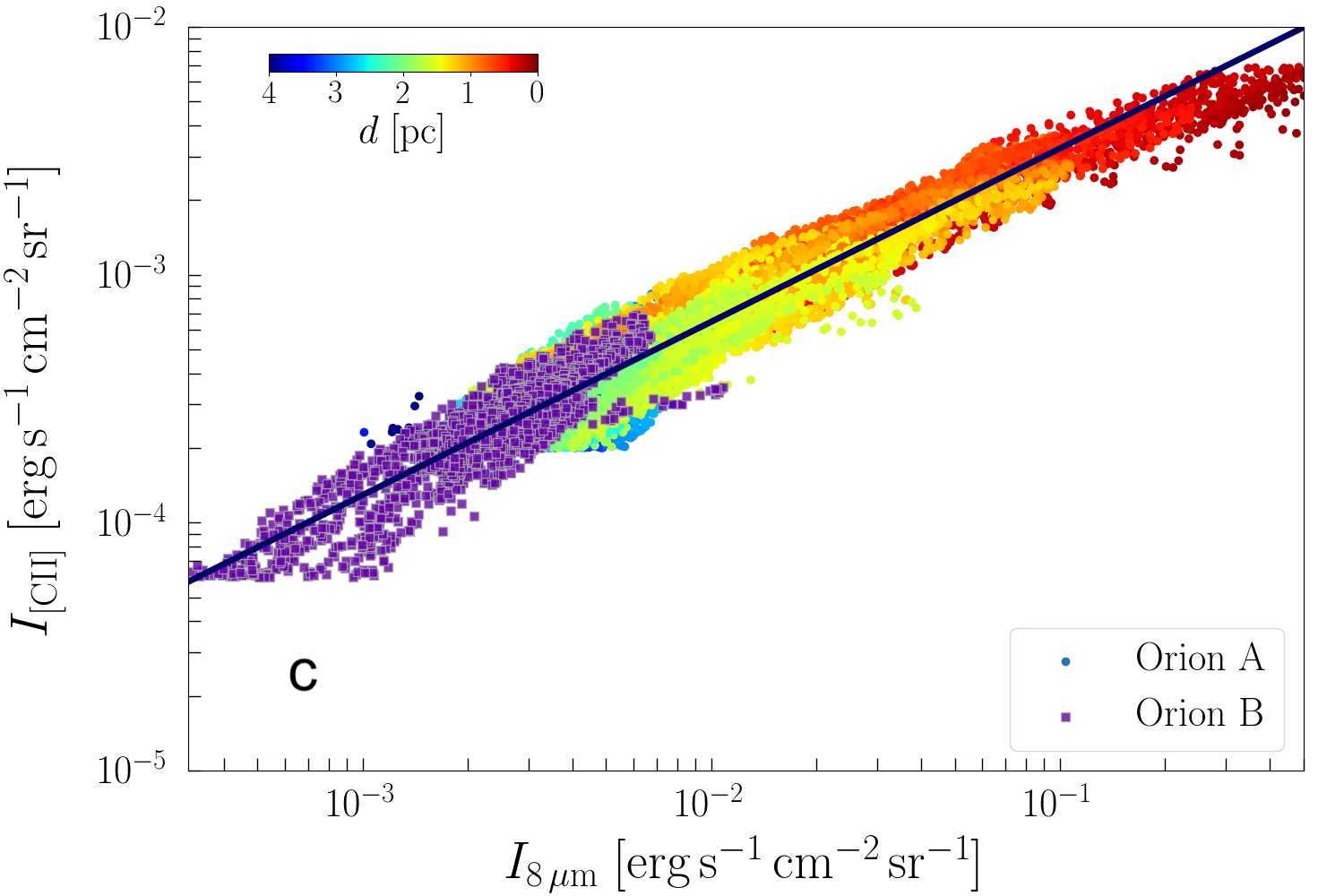}
\end{minipage}
\begin{minipage}{0.49\textwidth}
\includegraphics[width=\textwidth, height=0.67\textwidth]{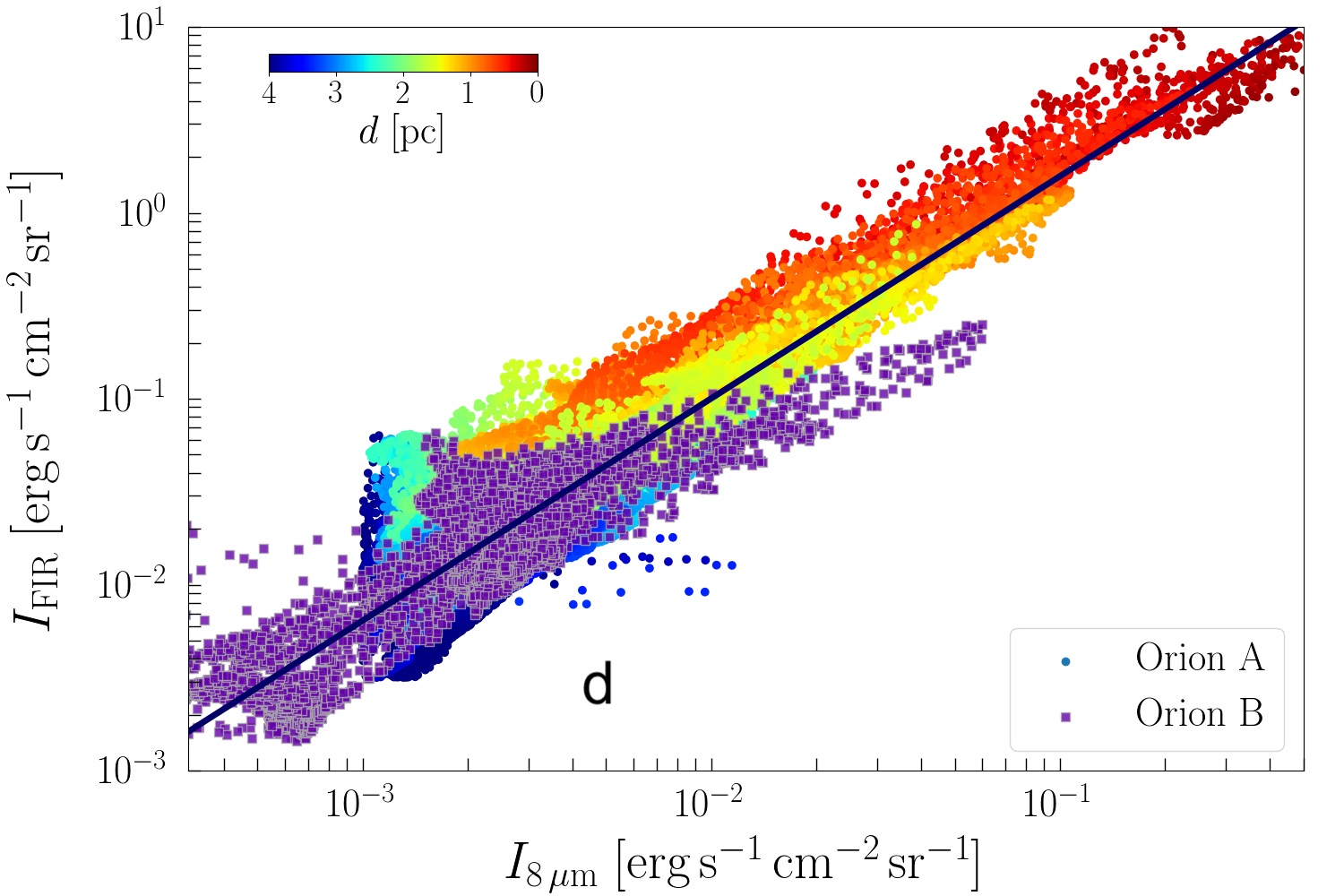}
\end{minipage}
\caption{Comparison of the correlations in Orion A and B: {\it a)} [C\,{\sc ii}]-$70\,\mu\mathrm{m}$, {\it b)} [C\,{\sc ii}]-FIR, {\it c)} [C\,{\sc ii}]-$8\,\mu\mathrm{m}$, {\it d)} FIR-$8\,\mu\mathrm{m}$. The color scale in Orion A indicates the distance from $\theta^1$ Ori C.}
\label{Fig.OrionB}
\end{figure*}

\begin{table}[ht]
\addtolength{\tabcolsep}{-2pt}
\def\arraystretch{1.2}
\caption{Summary of the density correlation plots in Fig. \ref{Fig.OrionB}.}
\begin{tabular}{ll|cccc}
\hline\hline
x & y & $a$ & $b$ & $\rho$ & $rms\,\mathrm{[dex]}$\\ \hline
$70\,\mu\mathrm{m}$ & {[C\,{\sc ii}]} & $0.50\pm 0.13$ & $-4.85\pm 0.39$ & 0.93 & 0.11\\
FIR & {[C\,{\sc ii}]} & $0.56\pm 0.15$ & $-2.65\pm 0.20$ & 0.94 & 0.11\\
$8\,\mu\mathrm{m}$ & {[C\,{\sc ii}]} & $0.70\pm 0.18$ & $-1.79\pm 0.40$ & 0.95 & 0.10\\
$8\,\mu\mathrm{m}$ & FIR & $1.19\pm 0.17$ & $1.39\pm 0.43$ & 0.93 & 0.20\\
\end{tabular}
\tablefoot{The regression fitted to the data is $\log_{10} y = a \log_{10} x + b$. Fit in $I_{\mathrm{[C\,\textsc{ii}]}}>3\sigma$ of the Orion B data. $\rho$ is the Pearson correlation coefficient, $rms$ is the root-mean-square of the residual of the fit.}
\label{Tab.fits_OrionB}
\end{table}

Figure \ref{Fig.OrionB} compares the correlations of the [C\,{\sc ii}] intensity with the $70\,\mu\mathrm{m}$, FIR, and $8\,\mu\mathrm{m}$ intensity, and the FIR-$8\,\mu\mathrm{m}$ correlation in Orion A and B. For Orion A, we plot the data in M42 only (with $I_{\mathrm{[C\,\textsc{ii}]}}>2\times 10^{-4}\,\mathrm{erg\,s^{-1}\,cm^{-2}\,sr^{-1}}$). Data of Orion B are taken from \cite{Pabst2017} and plotted for $I_{\mathrm{[C\,\textsc{ii}]}}>3\sigma\simeq 6\times 10^{-5}\,\mathrm{erg\,s^{-1}\,cm^{-2}\,sr^{-1}}$. The correlations we find from the combined data sets, summarized in Table \ref{Tab.fits_OrionB}, are very similar to those found from the Orion A data only (cf. Table \ref{Tab.fits_global}). The Orion A data form a perfect continuation of the earlier study of [C\,{\sc ii}] emission in Orion B, allowing us to extend the correlations over more than two orders of magnitude in [C\,{\sc ii}] intensity.

While Orion A is a region shaped by the stellar feedback from the locally formed massive stars, L1630 in Orion B is shaped by a chance encounter of a massive star with a molecular cloud. The O9.5V star $\sigma$ Ori in Orion B is approaching the cloud and photoevaporating the gas from its surface \citep{Ochsendorf2014}. On the other hand, the shell structures that emit most of the [C\,{\sc ii}] luminosities in Orion A are created by the stars in their respective centers. Regardless of the formation history of the emitting gas structures, the [C\,{\sc ii}] intensity is determined by PDR physics, as the tight correlations with the FIR and $8\,\mu\mathrm{m}$ intensity, respectively, reveal.

\subsection{The [C\,{\sc ii}] deficit}

Multiple studies of galaxies report that the [C\,{\sc ii}]/FIR ratio drops with increasing FIR luminosity and increasing FIR color temperature \citep[e.g.,][]{Malhotra2001,Luhman2003}. This is referred to as ``[C\,{\sc ii}] deficit''. Usually, other FIR lines are affected, as well \citep[e.g.,][]{Herrera-Camus2018b}. For the [C\,{\sc ii}] 158\,$\mu$m line, in particular, emission may be suppressed and cooling could come out in, for example, the [O\,{\sc i}] 63\,$\mu$m line (critical density of several $10^5\,\mathrm{cm^{-3}}$) in dense PDRs, the photoelectric heating rate depends non-linearly on the incident FUV radiation field as small dust grains and PAHs get charged at high radiation field, the relative contribution from different components of the ISM to the [C\,{\sc ii}] emission varies, and/or the FIR continuum emission has a large contribution from non-PDR dust. In (U)LIRGs, the [C\,{\sc ii}] deficit has been attributed to the importance of [O\,{\sc i}] cooling at high UV fields and densities as well as the importance of heating sources of dust other than stellar radiation \citep[e.g., AGN activity][]{Malhotra2001, Luhman2003, Croxall2012, Pineda2018, Rybak2020}. In more distant high-mass star-forming regions and in distant galaxies, some of the low [C\,{\sc ii}]/FIR ratios may be explained by [C\,{\sc ii}] absorption by foreground diffuse gas, for example diffuse clouds in the spiral arms \citep{Gerin2015}.
Nearby, well-resolved regions of massive star formation provide an ideal opportunity to study some of these effects in detail as any ``deficit''  can be linked to the characteristics of the source and the local physical conditions. The large-scale [C\,{\sc ii}] map of the Orion Nebula complex allows us to address some of the issues that may influence the [C\,{\sc ii}]/FIR ratio.

\begin{figure}[tb]
\includegraphics[width=0.5\textwidth, height=0.5\textwidth]{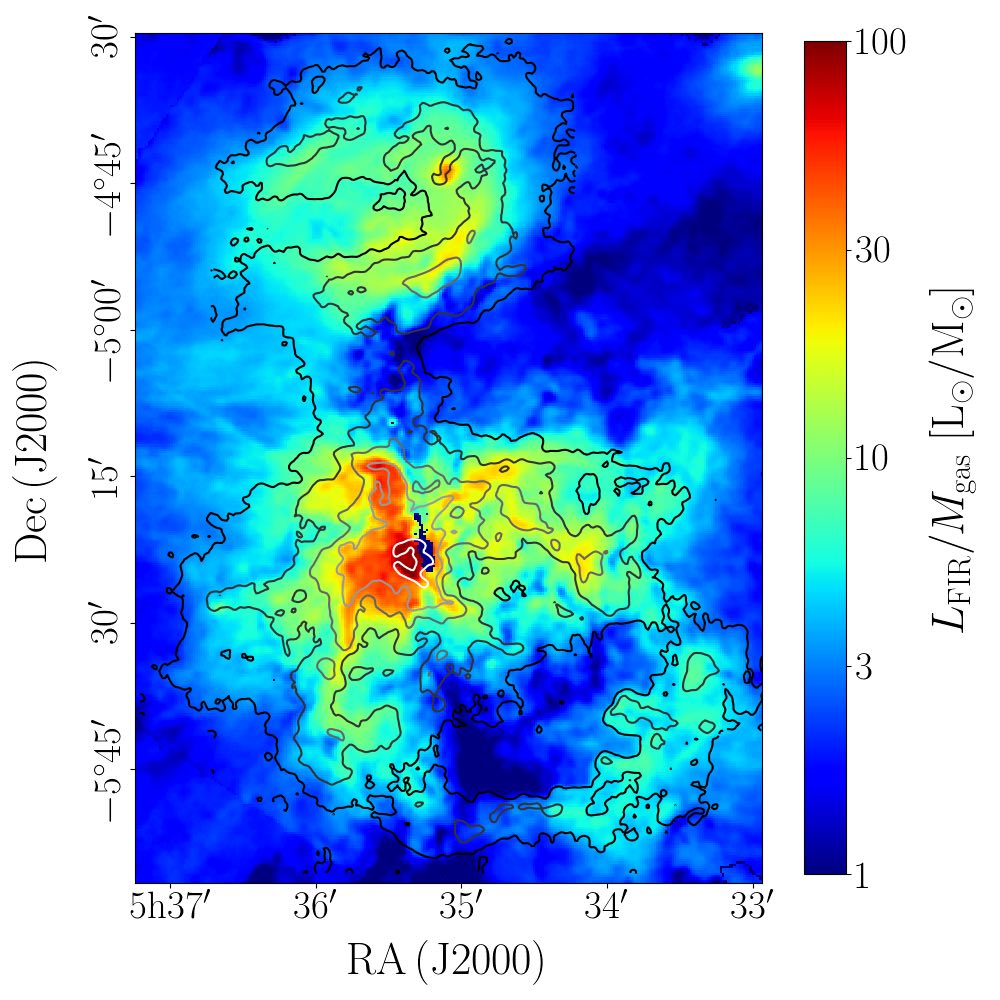}
\caption{Spatial distribution of $L_{\mathrm{FIR}}/M_{\mathrm{gas}}$ in Orion A with [C\,{\sc ii}] line-integrated intensity in contours (from black to white: $2, 5, 10, 20, 50\times 10^{-4}\,\mathrm{erg\,s^{-1}\,cm^{-2}\,sr^{-1}}$).}
\label{Fig.FIR-tau-ratio}
\end{figure}

Figure 1 in \cite{Herrera-Camus2018b} illustrates the [C\,{\sc ii}] deficit in a local sample of archetypical galaxies. The authors note that the threshold value in FIR surface luminosity, above which the [C\,{\sc ii}]/FIR ratio decreases, marks galaxies with $L_{\mathrm{FIR}}/M_{\mathrm{gas}}\gtrsim 50\,L_{\sun}/M_{\sun}$. Values of $L_{\mathrm{FIR}}/M_{\mathrm{gas}}\simeq 50\text{-}80\,L_{\sun}\,M_{\sun}^{-1}$ have been suggested to characterize galaxies with enhanced star formation \citep[e.g.,][]{Genzel2010}. In the Orion Nebula complex, high values of $L_{\mathrm{FIR}}/M_{\mathrm{gas}}$ (computed from the dust SEDs and using standard dust properties of \citet{Weingartner2001}) are found towards OMC1 and close to the central stars of M43 and NGC 1977 (see Fig. \ref{Fig.FIR-tau-ratio}). In fact, we can express the ratio as
\begin{align}
\frac{L_{\mathrm{FIR}}}{M_{\mathrm{gas}}}\simeq 4\pi\times 4\times 10^{-2} \int\limits_{40\,\mu\mathrm{m}}^{500\,\mu\mathrm{m}} B(\lambda,T_{\mathrm{d}})\left(\frac{160\,\mu\mathrm{m}}{\lambda}\right)^{\beta}\,\mathrm{d}\lambda \;\frac{L_{\sun}}{M_{\sun}}, \label{eq.L-M-ratio}
\end{align}
with $L_{\mathrm{FIR}}=4\pi A \tau_{160} \int\limits_{40\,\mu\mathrm{m}}^{500\,\mu\mathrm{m}} B(\lambda,T_{\mathrm{d}})(160\,\mu\mathrm{m}/\lambda)^{\beta}\,\mathrm{d}\lambda$ (in the optically thin limit) and $M_{\mathrm{gas}}=6\times 10^{24}\,\mathrm{cm^{-2}} \mu m_{\mathrm{H}}A \tau_{160} $, where $A$ is the surface area, $\mu=1.4$ the mean atomic weight, and $B(\lambda,T_{\mathrm{d}})\left(\frac{160\,\mu\mathrm{m}}{\lambda}\right)^{\beta}$ is the modified blackbody function (cf. Eq. \ref{eq.I}). In a second step, we approximate the integral over FIR wavelengths by the infinite integral $\int\limits_0^{\infty} B(\lambda,T_{\mathrm{d}})(160\,\mu\mathrm{m}/\lambda)^{\beta}\,\mathrm{d}\lambda \simeq 4.17\times 10^{-8}\,\mathrm{erg\,s^{-1}\,cm^{-2}\,sr^{-1}\,K^{-6}}\,T_{\mathrm{d}}^6$ for $\beta=2$, and obtain
\begin{align}
\frac{L_{\mathrm{FIR}}}{M_{\mathrm{gas}}} \simeq 2\times 10^{-2} \left(\frac{T_{\mathrm{d}}}{10\,\mathrm{K}}\right)^6 \;\frac{L_{\sun}}{M_{\sun}}. \label{eq.L-M-ratio-2}
\end{align}
This is accurate for optically thin dust FIR emission and $10\,\mathrm{K}\lesssim T_{\mathrm{d}}\lesssim 35\,\mathrm{K}$. For $T_{\mathrm{d}}=40\,\mathrm{K}$, the finite integral is overestimated by 10\%, and by 25\% for $T_{\mathrm{d}}=50\,\mathrm{K}$. Below $T_{\mathrm{d}}=10\,\mathrm{K}$, the dust blackbody emission rapidly shifts towards millimeter wavelengths and the infinite integral is not a good approximation for the FIR emission either\footnote{If the FIR intensity includes the wavelength range $500\text{-}1000\,\mu\mathrm{m}$, the infinite integral is accurate down to $T_{\mathrm{d}}\sim 5\,\mathrm{K}$.}.

Expressed in this way, the ratio $L_{\mathrm{FIR}}/M_{\mathrm{gas}}$ is independent of the gas column and depends solely on the {\it effective} dust temperature. Naturally, the effective dust temperature in Orion A is highest towards the ionized gas, that is close to the central heating source. We show a map of the dust temperature in Paper II. In our SED fits, the molecular ridge with the BN/KL region and Orion S is excluded \citep[for detailed SED fitting of this region, see][]{Chuss2019}. $L_{\mathrm{FIR}}/M_{\mathrm{gas}}$ assumes peak values of $80\,L_{\sun}\,M_{\sun}^{-1}$ in M42, $60\,L_{\sun}\,M_{\sun}^{-1}$ in M43, and $20\,L_{\sun}\,M_{\sun}^{-1}$ in NGC 1977 (with the exception of $40\,L_{\sun}\,M_{\sun}^{-1}$ towards NGC 1973). These highest values are concentrated in the H\,{\sc ii} regions of M43 and NGC 1977, like in OMC1. From SED fits of \cite{Chuss2019}, we learn that exceedingly high values, $L_{\mathrm{FIR}}/M_{\mathrm{gas}}\sim 10^3\,L_{\sun}\,M_{\sun}^{-1}$, are found in a small region around the BN/KL core with $T_{\mathrm{d}}\simeq 95\,\mathrm{K}$. The global $L_{\mathrm{FIR}}/M_{\mathrm{gas}}$ ratio for M42 amounts to $14\,L_{\sun}\,M_{\sun}^{-1}$, while in M43 it is $26\,L_{\sun}\,M_{\sun}^{-1}$, and $7\,L_{\sun}\,M_{\sun}^{-1}$ in NGC 1977. Thus, globally, the Orion Nebula is characterized by rather low $L_{\mathrm{FIR}}/M_{\mathrm{gas}}$ and probably not representative of environments that dominate the FIR emission in ULIRGs, that may comprise large amounts of embedded star-forming cores like BN/KL. As we will discuss below, those embedded sources show a FIR excess in the continuum emission compared to the line emission.

In conclusion, in particular dense H\,{\sc ii} regions (like the Huygens Region and M43) exhibit high $L_{\mathrm{FIR}}/M_{\mathrm{gas}}$ ratios in their centers. $L_{\mathrm{FIR}}/M_{\mathrm{gas}}$ is expected to be proportional to the H\,{\sc ii} ionization parameter $U = Q(H)/4\pi R^2/c/n_e$ \citep{Abel2009}. Indeed, $U$ is similar in M43 and NGC 1977 ($U\simeq 10^{-3}$), but higher in the Huygens Region ($U\simeq 10^{-2}$). The $L_{\mathrm{FIR}}/M_{\mathrm{gas}}$ ratio in Orion is dominated by the FIR emission from the dust column of the limb-brightened edges and the background molecular cloud (in M42 and M43) or the enveloping expanding shell (in NGC 1977) and does not trace the ionized gas directly. Yet, the properties of a PDR are closely related to the properties of the central H\,{\sc ii} region \citep{YoungOwl2002,Abel2005,Seo2019} and discussed further in Paper II.

\begin{figure*}[tb]
\begin{minipage}[t]{0.49\textwidth}
\includegraphics[width=\textwidth, height=0.67\textwidth]{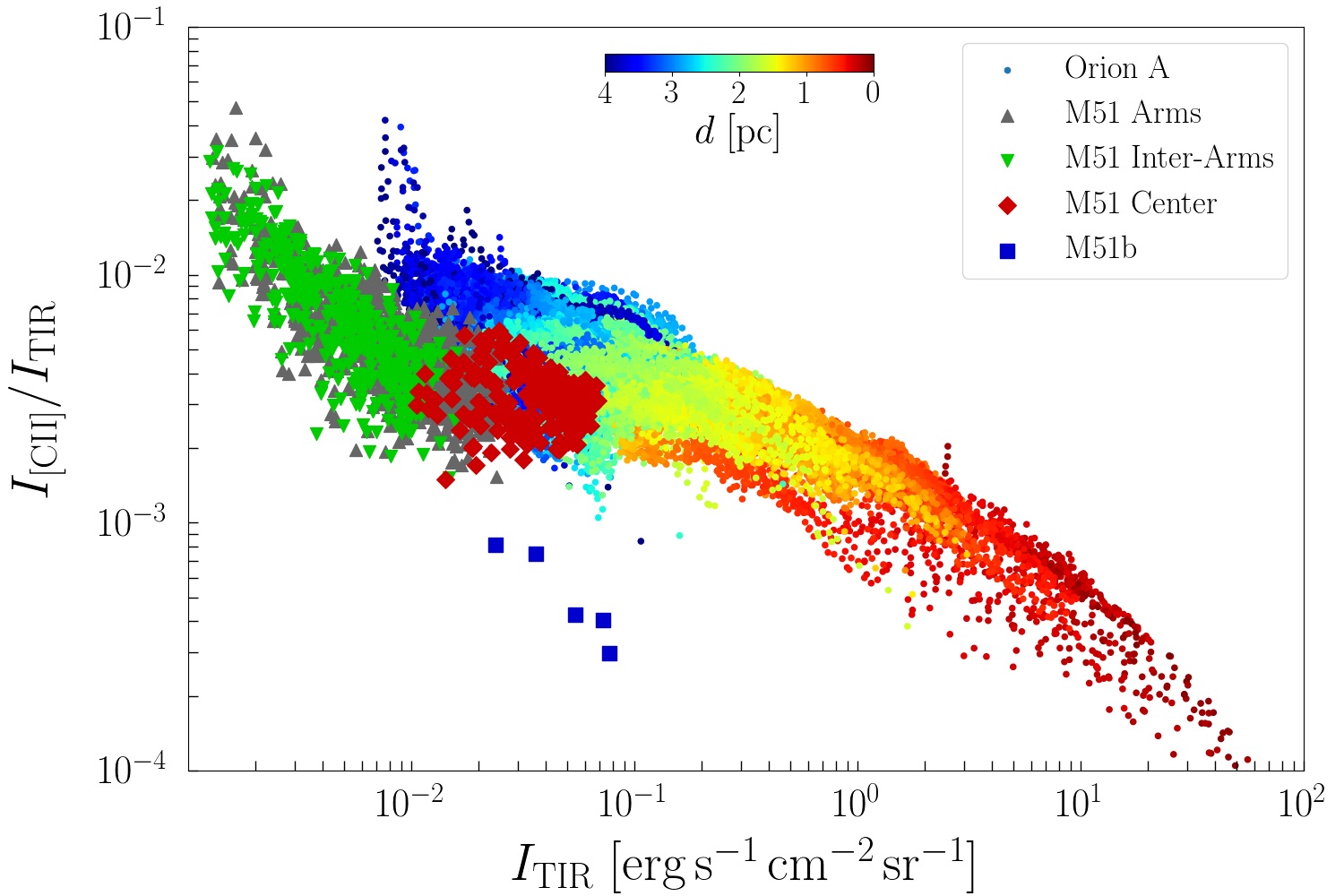}
\end{minipage}
\begin{minipage}[t]{0.49\textwidth}
\includegraphics[width=\textwidth, height=0.67\textwidth]{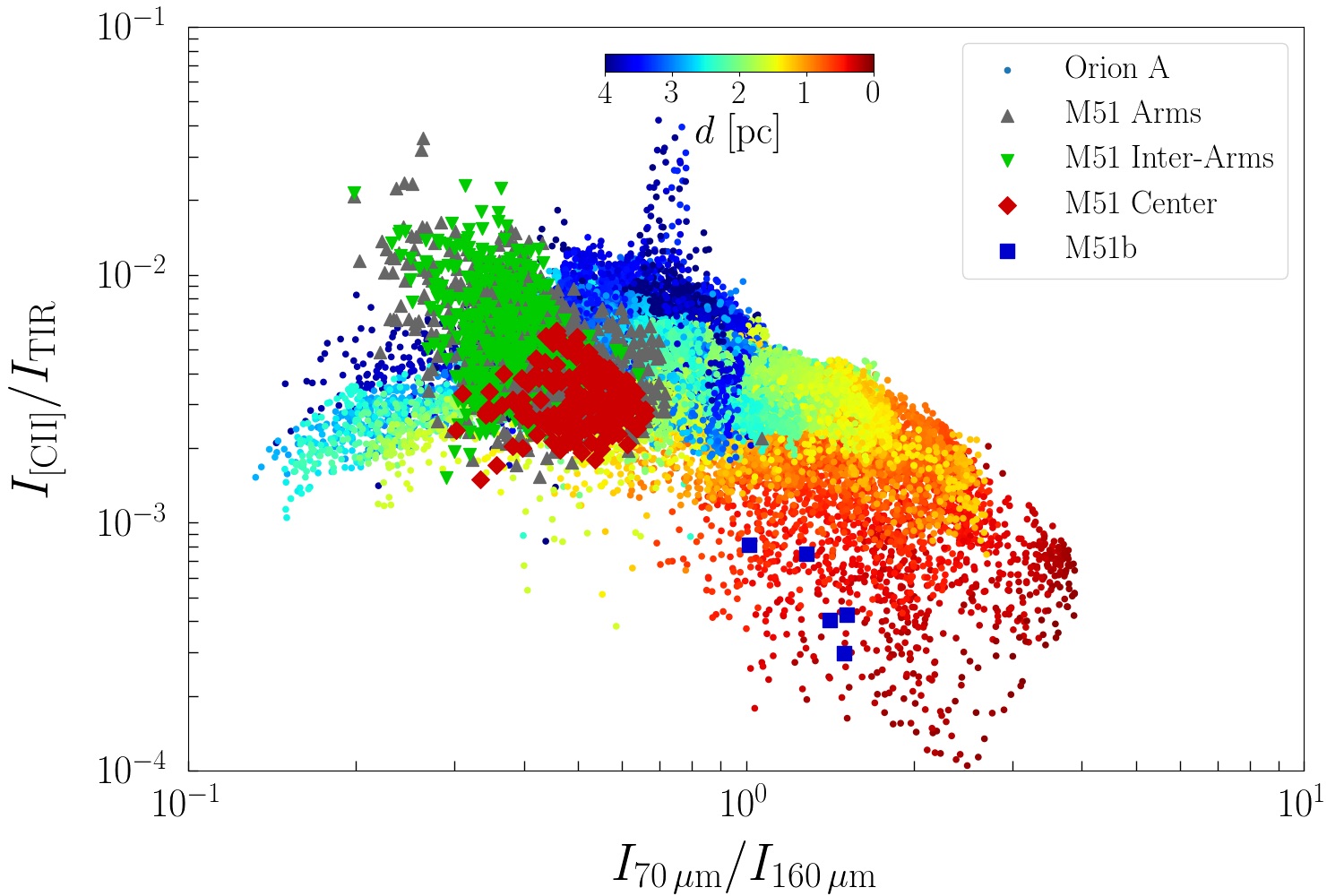}
\end{minipage}
\caption{{\it Left}: [C\,{\sc ii}]/TIR intensity ratio versus TIR intensity in M51 and the Orion Nebula with M43. {\it Right}: [C\,{\sc ii}]/TIR intensity ratio versus PACS $70\,\mu\mathrm{m}$ over $160\,\mu\mathrm{m}$ intensity ratio in M51 and the Orion Nebula with M43. Fig. 1 in \cite{Pineda2018} shows the mask used to divide the M51 map into the four separate regions displayed here. Colors in Orion indicate distance from Trapezium stars as before (see colorbar).}
\label{Fig.M51-CII-TIR}
\end{figure*}

We now turn to discuss the [C\,{\sc ii}] line deficit in spatially resolved samples, Orion A and M51. Figure 2 of \cite{Pineda2018} shows the [C\,{\sc ii}]/TIR versus TIR correlation and the [C\,{\sc ii}]/TIR versus $70\,\mu\mathrm{m}$/$160\,\mu\mathrm{m}$ correlation in the spiral galaxy M51 (with a spatial resolution of $16\arcsec$ corresponding to $660\,\mathrm{pc}$). Following \cite{Pineda2018}, we compute the total infrared \citep[TIR, integrated between 3 and $1000\,\mu\mathrm{m}$;][]{Croxall2012} intensity by
\begin{align}
I_{\mathrm{TIR}} = 0.95\nu I_{\nu,8\,\mu\mathrm{m}} + 1.15\nu I_{\nu,24\,\mu\mathrm{m}} + \nu I_{\nu,70\,\mu\mathrm{m}} + \nu I_{\nu,160\,\mu\mathrm{m}}.
\end{align}
Since the MIPS $24\,\mu\mathrm{m}$ image is saturated towards OMC1, we approximate $I_{\nu,24\,\mu\mathrm{m}}\simeq 0.1 I_{\nu,70\,\mu\mathrm{m}}$, which is valid for most of the bright, edge-on structures in the Orion Nebula (see Paper II). With this, we find $I_{\mathrm{TIR}}/I_{\mathrm{FIR}} \simeq 2$ towards the Orion Nebula. As extragalactic observations of regions of massive star formation are generally beam diluted, it is better to compare the relation between the [C\,{\sc ii}]/FIR and the $70\,\mu\mathrm{m}$-$160\,\mu\mathrm{m}$ color temperature. This latter ratio is a measure for the dust temperature. We show both correlations in Fig. \ref{Fig.M51-CII-TIR}, where for Orion we plot only points associated with M42 and M43. Points in NGC 1977 follow the same trend, but cover only the low-TIR end of the correlation. The trend of \cite{Pineda2018} is very similar to the trend observed in the Orion Nebula complex. The M51 relationship is slightly offset from the Orion data to lower $70\,\mu\mathrm{m}$/$160\,\mu\mathrm{m}$ ratios and/or lower [C\,{\sc ii}]/FIR. We ascribe this to a contribution of diffuse neutral HI clouds in the interstellar medium of M51 which is characterized by cooler dust.

In Orion, the [C\,{\sc ii}]/FIR ratio drops particularly for regions that are close to the illuminating star. These are regions with $G_0>10^4$ and $n>4\times 10^4\,\mathrm{cm^{-3}}$. For these regions cooling by [O\,{\sc i}] $63\,\mu\mathrm{m}$ is known to be important with $I_{\mathrm{[O\,\textsc{i}]\,63\mu m}}/I_{\mathrm{[C\,\textsc{ii}]\,158\mu m}}\simeq 5\text{-}10$ \citep{Herrmann1997}. Following \cite{Luhman2003} and \cite{Croxall2012}, we attribute the drop in the [C\,{\sc ii}]/FIR relation with increasing FIR or $70\,\mu\mathrm{m}$/$160\,\mu\mathrm{m}$ color temperature to a shift of the gas cooling to [O\,{\sc i}] $63\,\mu\mathrm{m}$ line. Implicitly, this assumes that the heating efficiency does not depend strongly on the physical conditions in the emitting region. Variations in the heating efficiency are generally ascribed to variations in the ionization balance, which is controlled by the PDR ionization parameter $\gamma=G_0 T^{1/2}/n_{\rm e}$ \citep[cf. Section \ref{Sec.pe-efficiency};][]{BakesTielens1994}. The observed relation between $G_0$ and the thermal pressure $p_{\mathrm{th}}=nk_{\mathrm{B}}T$ (see Paper II) translates this into a very weak dependence of the ionization parameter on physical conditions. We note that the [C\,{\sc ii}] optical depth of the bright [C\,{\sc ii}]-emitting structures in Orion A is usually of the order $\tau_{\mathrm{[C\,\textsc{ii}]}}\simeq 1\text{-}3$, both in the edge-on shells and in the very bright core OMC1 \citep[see Paper II;][]{Goicoechea2015}, which may affect the correlations slightly, but we deem the effect too small to correct for the [C\,{\sc ii}] deficit. When estimating star-formation rates of galaxies, however, the [C\,{\sc ii}] optical depth has to be taken into account, as \cite{Okada2019} point out.

\begin{figure}[tb]
\includegraphics[width=0.5\textwidth, height=0.33\textwidth]{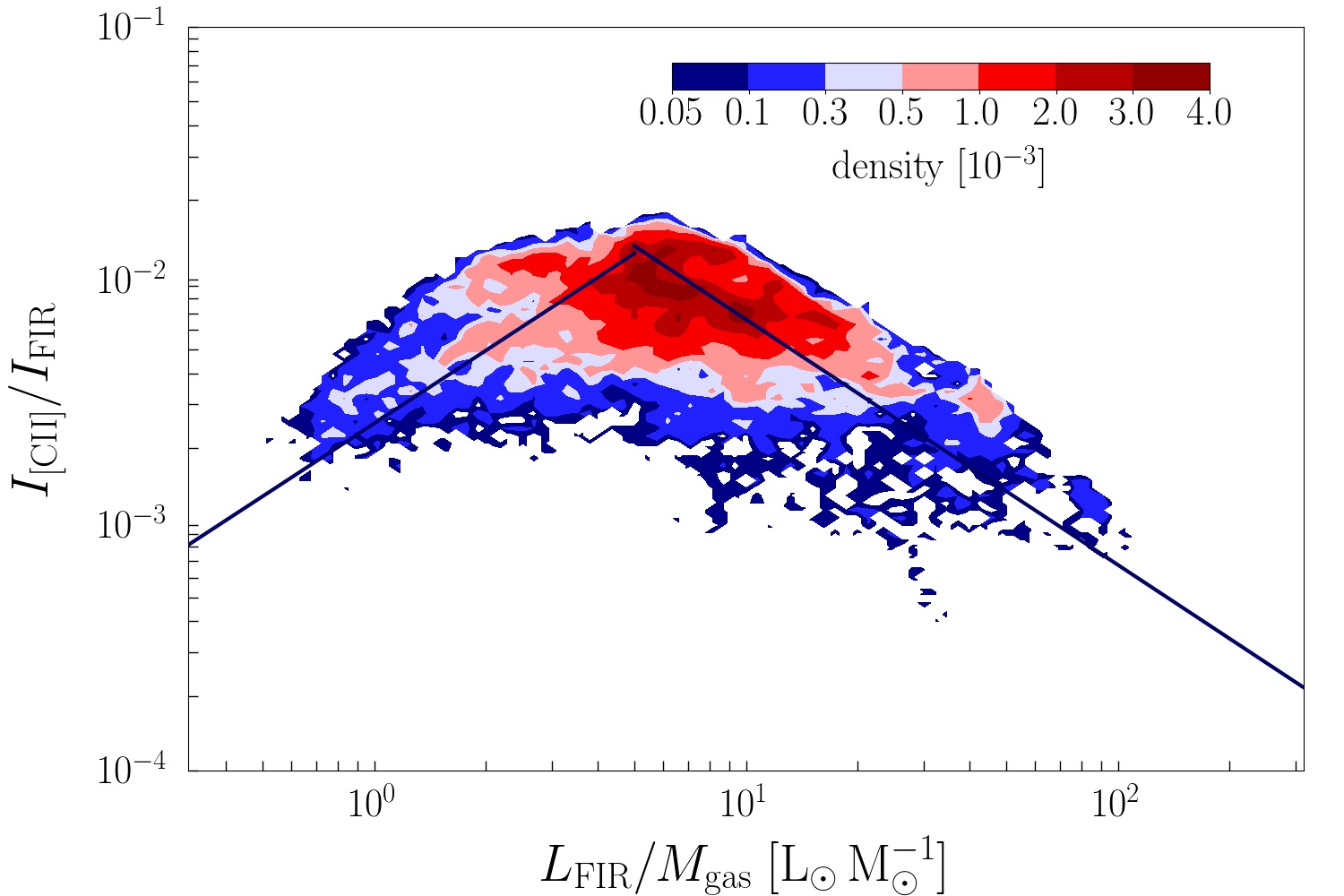}
\caption{[C\,{\sc ii}]/FIR intensity ratio versus $L_{\mathrm{FIR}}/M_{\mathrm{gas}}$ ratio. Fit for $L_{\mathrm{FIR}}/M_{\mathrm{gas}}>5$: $\log_{10} I_{\mathrm{[C\,\textsc{ii}]}}/I_{\mathrm{FIR}} \simeq (-0.99\pm 0.86) \log_{10} L_{\mathrm{FIR}}/M_{\mathrm{gas}} (-1.19\pm 0.87)$, $rms\simeq 0.21\,\mathrm{dex}$, $\rho\simeq -0.65$; fit for $L_{\mathrm{FIR}}/M_{\mathrm{gas}}<5$: $\log_{10} I_{\mathrm{[C\,\textsc{ii}]}}/I_{\mathrm{FIR}} \simeq (0.98\pm 0.62) \log_{10} L_{\mathrm{FIR}}/M_{\mathrm{gas}} (-2.60\pm 0.27)$, $rms\simeq 0.20\,\mathrm{dex}$, $\rho\simeq 0.53$.}
\label{Fig.CII-FIR-M}
\end{figure}

Besides close to the central heating sources, where [O\,{\sc i}] cooling is dominant, low [C\,{\sc ii}]/FIR ratios ($I_{\mathrm{[C\,\textsc{ii}]}}/I_{\mathrm{FIR}} \simeq 2\times 10^{-3}$) are found along the dense spine of the ISF (cf. Fig. \ref{Fig.contours}f). Considering the low temperature of these structures, we do not expect [O\,{\sc i}] emission to contribute significantly. Rather, we interpret these ratios as a ``FIR excess'' from cold dust in the dense molecular gas. Particularly deficient [C\,{\sc ii}]/FIR ratios ($I_{\mathrm{[C\,\textsc{ii}]}}/I_{\mathrm{FIR}} \simeq 2\times 10^{-4}$) are also found towards the active star-forming cores in the BN/KL region and Orion S \citep[FIR intensity from][]{Goicoechea2015b}. Here, [O\,{\sc i}] cooling is dominant \citep[$I_{\mathrm{[O\,\textsc{i}]\,63\mu m}}/I_{\mathrm{[C\,\textsc{ii}]\,158\mu m}}\simeq 20$][]{Herrmann1997}, but in addition to excitation by FUV radiation at the surface, the [O\,{\sc i}] lines have a contribution from shocked gas in the interior \citep{Goicoechea2015b}. Also the FIR emission stems from the inner layers of embedded star formation. As opposed to the FIR excess from cold dust in the ISF, this FIR excess is produced by warm molecular gas in the star-forming cores. The heavily irradiated PDR on the surface of the molecular core east of the Trapezium stars, devoid of star-forming activity, shows less deficient FIR line ratios, $(I_{\mathrm{[O\,\textsc{i}]}}+I_{\mathrm{[C\,\textsc{ii}]}})/I_{\mathrm{FIR}}\simeq 5\text{-}10\times 10^{-3}$. Large columns of warm dust, such as in BN/KL and Orion S, lead automatically to high $L_{\mathrm{FIR}}/M_{\mathrm{gas}}$ ratios, as we have shown by Eq. \ref{eq.L-M-ratio}, whereas the effective dust temperature will be lower for an irradiated surface with a large column of cooler (not internally heated) material behind it. Fig. \ref{Fig.CII-FIR-M} shows that the [C\,{\sc ii}]/FIR ratio in Orion A decreases with increasing $L_{\mathrm{FIR}}/M_{\mathrm{gas}}$ ratio for $L_{\mathrm{FIR}}/M_{\mathrm{gas}}\gtrsim 5$. At lower $L_{\mathrm{FIR}}/M_{\mathrm{gas}}$ ratios, however, the [C\,{\sc ii}]/FIR ratio tends to increase with increasing $L_{\mathrm{FIR}}/M_{\mathrm{gas}}$.

In Section \ref{Sec.correlations}, we have shown that the [C\,{\sc ii}] luminosity depends less than linear on the FIR luminosity per pixel. While the regions we study have different stellar content, the [C\,{\sc ii}]-FIR correlation is very similar in all of the regions. Moreover, we note that the [C\,{\sc ii}]/FIR versus FIR in the L1630 region \citep{Pabst2017} is a good continuation of the correlation in the central OMC1 \citep{Goicoechea2015}. Yet, L1630 represents the chance encounter of an O star with a molecular cloud \citep{Ochsendorf2014} rather than emission from a region of active massive star formation. This provides further support for an interpretation of the [C\,{\sc ii}]/FIR relation in terms of the physical conditions rather than the star-formation history of the region. On the other hand, we have argued that particularly low FIR line-to-continuum ratios are found towards regions of large columns of (warm and cold) dust, which suggests the interpretation of the FIR line ``deficit'' in terms of a ``FIR excess''.

\subsection{The photoelectric heating efficiency}
\label{Sec.pe-efficiency}

\begin{figure*}[tb]
\begin{minipage}{0.49\textwidth}
\includegraphics[width=\textwidth, height=0.67\textwidth]{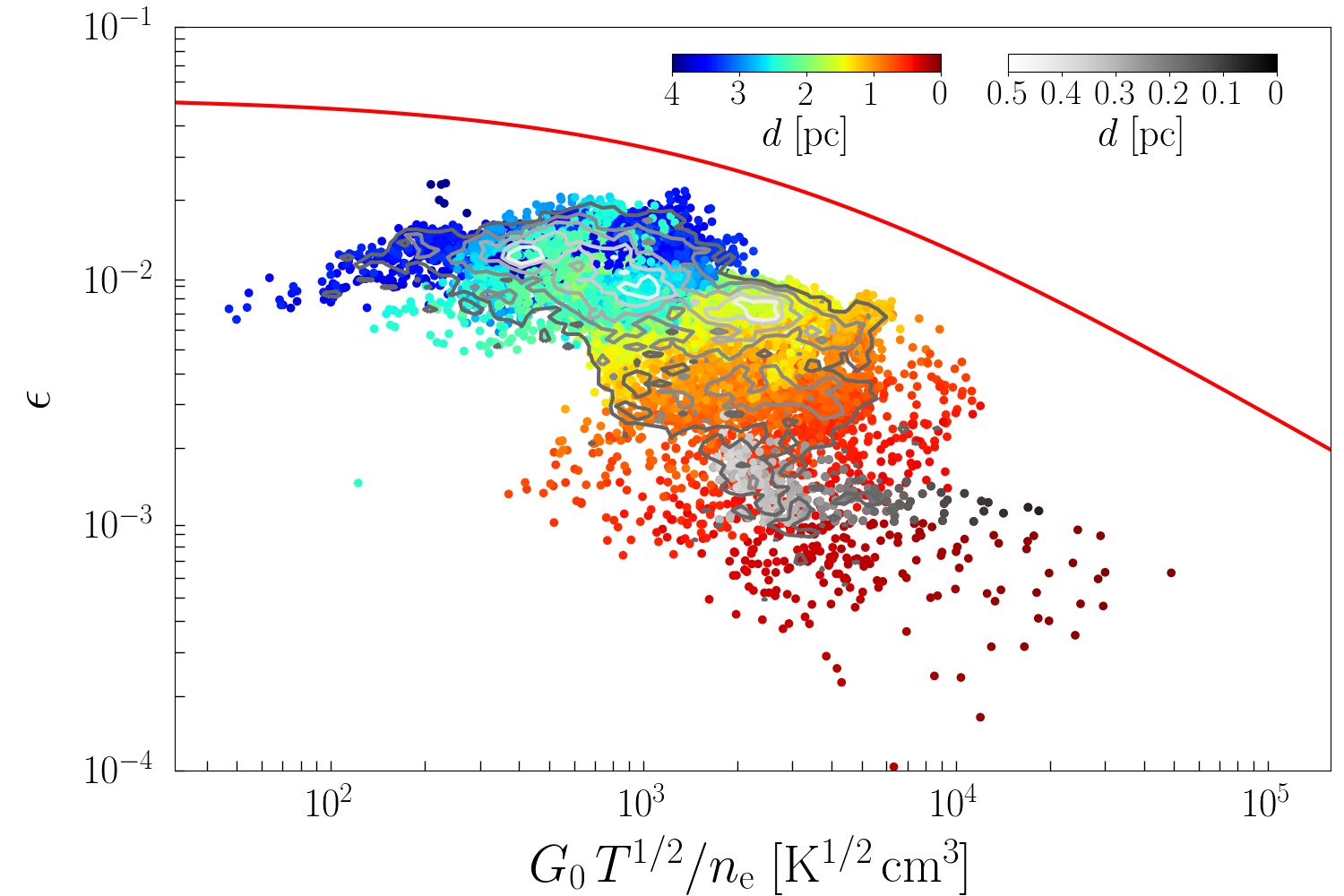}
\end{minipage}
\begin{minipage}{0.49\textwidth}
\includegraphics[width=\textwidth, height=0.67\textwidth]{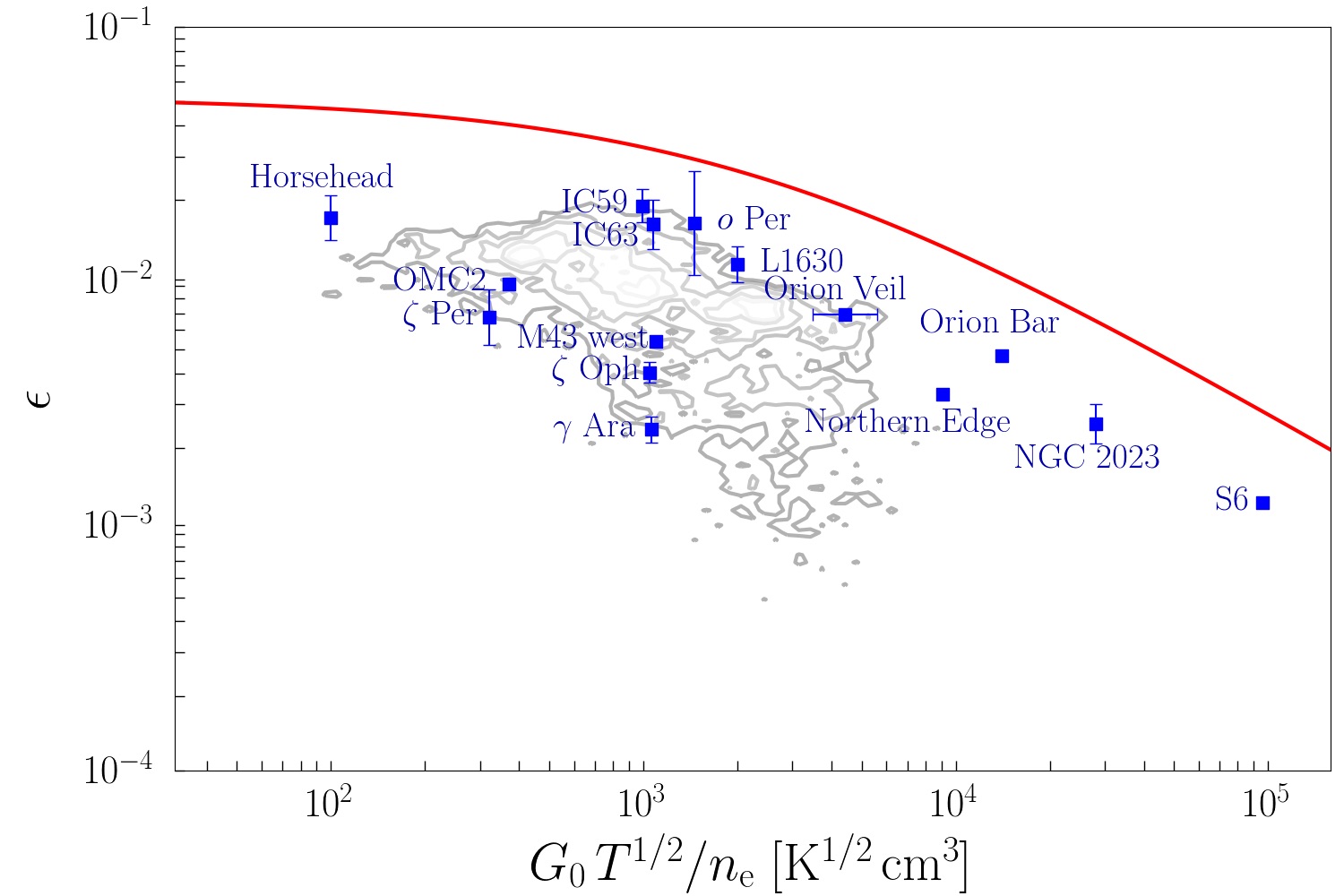}
\end{minipage}
\caption{{\it Left:} The photoelectric heating efficiency, equated as cooling efficiency [C\,{\sc ii}]/FIR, throughout the Orion Nebula as function of the ionization parameter $\gamma=G_0 T^{1/2}/n_{\mathrm{e}}$. Color scales indicate the distance from $\theta^1$ Ori C. Contours indicate the density levels above which 90, 70, 50, 30 and 10\% (from light to dark grey) of the points lie. The red curve is the theoretical prediction of Eq. 43 by \cite{BakesTielens1994}. {\it Right:} The photoelectric heating efficiency $\epsilon$ in PDRs and diffuse sight lines, equated as the cooling efficiency by [C\,{\sc ii}] and [O\,{\sc i}] cooling, overlaid on contours from the left-hand panel. Data for the diffuse ISM are taken from \citet{Gry1992} and \citet{vanDishoeck1986}, data of NGC 2023 are from \citet{HollenbachTielens1999}, data for the Horsehead and L1630 are from \citet{Pabst2017}, data for IC 59 and IC 63 are from \citet{Andrews2018}, data for the Orion Bar, S6, M43 west, and OMC2 are from \citet{Herrmann1997}, and data for the Orion Veil are from \citet{Salas2019}.}
\label{Fig.pe_scatter}
\end{figure*}

The heating mechanism of the interstellar gas is of great importance for the understanding of the properties of the ISM. The UV-illuminated gas is dominantly heated by the photoelectric effect acting on small grains and PAHs \citep{BakesTielens1994,Wolfire1995}. The efficiency of this process is a crucial parameter in models of the ISM. Assuming equilibrium of heating and cooling processes, we can estimate the photoelectric heating efficiency by the cooling efficiency of the gas. At intermediate temperatures and densities the cooling is dominated by cooling through the [C\,{\sc ii}] line. At higher temperatures and densities, like those in the Huygens Region, the [O\,{\sc i}] $63\,\mu\mathrm{m}$ line becomes important \citep{HollenbachTielens1999}. However, in the shell surrounding the Extended Orion Nebula, gas cooling is most likely dominated by [C\,{\sc ii}], as the temperatures and densities are rather moderate. In M43, [O\,{\sc i}] $63\,\mu\mathrm{m}$ emission will be important, in NGC 1977 only in the densest part of the shell abutting OMC3 \citep{Herrmann1997,YoungOwl2002}.

Theoretically, the photoelectric heating efficiency depends on the ionization parameter $\gamma=G_0 T_{\mathrm{gas}}^{1/2}/n_{\mathrm{e}}$ \citep[Eq. 43]{BakesTielens1994}. We equate the heating efficiency in our sample with the [C\,{\sc ii}] cooling efficiency $I_{\mathrm{[C\,\textsc{ii}]}}/I_{\mathrm{FIR}}$, which is accurate for most points in our data set. In App. \ref{Sec.gamma}, we estimate $\gamma$ in each pixel of the map. The correlation between the photoelectric heating efficiency $\epsilon$ and the ionization parameter $\gamma$ in the Orion Nebula is shown in Fig. \ref{Fig.pe_scatter}. Points in NGC 1977 overlap and are not shown. The data reveal a similar trend with a decreased efficiency with increased ionization parameter as the theoretical predictions of \cite{BakesTielens1994}. The offset between the two might reflect a somewhat lower abundance of PAHs and VSGs in Orion than assumed by \cite{BakesTielens1994}. Closer to the central source, [O\,{\sc i}] cooling dominates. Adding the [O\,{\sc i}] $63\,\mu\mathrm{m}$ intensity to the [C\,{\sc ii}] intensity, shifts the heating/cooling efficiency to $\epsilon\simeq 7\times 10^{-3}$ (outside of BN/KL) and corrects for the steep decline in the [C\,{\sc ii}]-TIR plot (with $I_{\mathrm{[O\,\textsc{i}]}}/I_{\mathrm{[C\,\textsc{ii}]}}\simeq 6$ in OMC1). Nevertheless, the heating efficiency seems to slightly decline with radiation field, from $\sim 10^{-2}$ in the distant shell to $5\times 10^{-3}$ in the very center. While the [C\,{\sc ii}] deficit (and other FIR line deficits) observed on galactic scales might reflect different physical processes (XDRs, AGN activity), the [C\,{\sc ii}] deficit in the Orion Nebula reflects PDR physics with [O\,{\sc i}] cooling correcting for the majority of the [C\,{\sc ii}] deficit and a slightly decreased heating efficiency at higher radiation field causing the slight decline.

\begin{table*}[tb]
\centering
\addtolength{\tabcolsep}{-1.5pt}
\def\arraystretch{1.2}
\caption{Luminosities and masses of M42, M43, and NGC 1977.}
\begin{tabular}{llccccccccc}
\hline\hline
Region & & $A$ & $L_{\mathrm{FIR}}$ & $L_{\mathrm{[C\,\textsc{ii}]}}$ & $L_{\mathrm{[C\,\textsc{ii}]}}/L_{\mathrm{FIR}}$ & $M_{\mathrm{gas}}$\tablefootmark{a} & $L_{\mathrm{CO(2\text{-}1)}}$\tablefootmark{b} & $M_{\mathrm{CO}}$\tablefootmark{c} & $L_{\mathrm{24\mu m}}$\tablefootmark{d} & $L_{\mathrm{H\alpha}}$\tablefootmark{e} \\
&  &  $\mathrm{[pc^2]}$ & $\mathrm{[L_{\sun}]}$ & $\mathrm{[L_{\sun}]}$ & & $\mathrm{[M_{\sun}]}$ & $\mathrm{[L_{\sun}]}$ & $\mathrm{[M_{\sun}]}$ & $\mathrm{[L_{\sun}]}$ & $\mathrm{[L_{\sun}]}$ \\ \hline
M42 & all & 45 & $1.8\times 10^5$\tablefootmark{f} & 510 & $2.8\times 10^{-3}$ & 10000 & 1.1 & 8800 & 4700 & 3500 \\
 & OMC1 & 0.4 & $8.2\times 10^4$\tablefootmark{f} & 53 & $6.5\times 10^{-4}$ & 640 & $8.9\times 10^{-2}$ & 720 & -- & 1500 \\
 & shell & 10.0 & $5.2\times 10^4$ & 209 & $4.0\times 10^{-3}$ & 3000 & $3.2\times 10^{-1}$ & 2500 & 2000 & 850 \\

M43 & all & 0.74 & $1.4\times 10^4$ & 38 & $2.7\times 10^{-3}$ & 530 & $3.0\times 10^{-2}$ & 240 & 740 & 190 \\
 & background, H\,{\sc ii} & 0.13 & $3.1\times 10^3$ & 8.9\tablefootmark{f} & $2.8\times 10^{-3}$ & 53 & $4.6\times 10^{-3}$ & 37 & 100 & 99 \\
 & shell & 0.31 & $8.2\times 10^3$ & 19 & $2.3\times 10^{-3}$ & 250 & $1.4\times 10^{-2}$ & 110 & 410 & 63 \\

NGC 1977 & all & 17 & $1.8\times 10^4$ & 160 & $8.9\times 10^{-3}$ & 2600 & $1.2\times 10^{-1}$ & 950 & 1000 & 63 \\
 & OMC3 & 2.6 & $5.8\times 10^3$ & 34 & $5.9\times 10^{-3}$ & 1200 & $8.9\times 10^{-2}$ & 720 & 190 & 8.6 \\
 & shell & 6.2 & $5.4\times 10^3$ & 69 & $1.3\times 10^{-2}$ & 670 & $1.6\times 10^{-2}$ & 130 & 100 & 7.9 \\ \hline
\end{tabular}
\tablefoot{\tablefoottext{a}{Gas mass derived from dust mass, using $N_H\simeq 6\times 10^{24}\,\mathrm{cm^{-2}}\,\tau_{160}$ \citep{Weingartner2001}.}
\tablefoottext{b}{ The CO(2-1) map does not cover the entire NGC 1977 region (cf. Fig. \ref{Fig.contours}).}
\tablefoottext{c}{Molecular gas mass from CO(2-1) intensity, using $X(\mathrm{CO})\simeq 2\times 10^{20}\,\mathrm{cm^{-2}}\,(\mathrm{K\,km\,s^{-1}})^{-1}$ \citep{Bolatto2013}.}
\tablefoottext{d}{The MIPS $24\,\mu\mathrm{m}$ image is saturated towards the Huygens/OMC1 region.}
\tablefoottext{e}{The H$\alpha$ surface brightness in the EON is largely due to scattered light from the bright Huygens Region \citep{ODell2010}. Also H$\alpha$ emission in M43 has to be corrected for a contribution of scattered light from the Huygens Region \citep{Simon-Diaz2011}.}
\tablefoottext{f}{BN/KL and Orion S contribute with $L_{\mathrm{FIR}}\simeq 3\times 10^4\,L_{\sun}$.}}
\label{Tab.Luminosities}
\end{table*}

\subsection{The origin of [C\,{\sc ii}] emission}

Table \ref{Tab.Luminosities} gives the luminosities and masses of M42, M43, and NGC 1977, total and for the respective brightest region and the extended shell structure. Paper II discusses the origin of the [C\,{\sc ii}] emission in greater detail.

Most of the [C\,{\sc ii}] emission stems from the limb-brightened shell edges in M42, M43, and NGC 1977 (30-50\%), but a significant contribution also comes from the more diffuse emission (25-45\%), that is faint emission from extended surfaces. Even though the regions close to the central stars are very bright, their area is small and the contribution to the total [C\,{\sc ii}] emission minor (10-20\%) compared to the large shells. For FIR emission, 20-30\% stem from the shells, 20-30\% from the regions close to the central stars of M43 and NGC 1977, but 45\% from OMC1 in M42, and 10-20\% from diffuse emission.

In Galactic samples, \cite{Pineda2013} note that most of the [C\,{\sc ii}] emission arises in moderately FUV-illuminated regions ($G_0\simeq 2\text{-}50$ in Habing units), that is large faint surfaces rather than small extreme regions. Our observations towards the Orion Nebula are reminiscent of this, the bright inner OMC1 region being a minor contributor to the total [C\,{\sc ii}] luminosity. The [C\,{\sc ii}] observations towards L1630, comprising the Horsehead Nebula, predominantly highlight the PDR surfaces illuminated by $\sigma$ Ori, where 95\% of the total [C\,{\sc ii}] emission in the mapped area arise. Yet, regions that are bright in [C\,{\sc ii}] trace only 8\% of the gas mass, while 85\% of the mass is associated with strong CO emission \citep{Pabst2017}. In the Orion Nebula complex, we find that 70\% of the [C\,{\sc ii}] emission arises in bright PDR surfaces with $G_0\gtrsim 100$. The [C\,{\sc ii}]-emitting regions contain 64\% of the gas mass (traced by dust). A portion of that gas mass, about 30-50\%, is likely associated with the molecular background rather than the PDR surface. Hence, bright [C\,{\sc ii}] emission, that traces the H/H$_2$ transition in a PDR, traces about 30\% of the total gas mass in the Orion Nebula complex.

We conclude that [C\,{\sc ii}] emission from PDRs with $G_0\gtrsim 100$ is the main origin of [C\,{\sc ii}] emission from the Orion Nebula complex, with about 70\%. The [C\,{\sc ii}] emission from ionized gas is a minor contributor, less than 5\% in M42 and NGC 1977, but 15\% in M43. From the remainder, we estimate that less irradiated gas that is not captured in bright PDR surfaces, contributes with about 20\% within the mapped area. These percentages elucidate the importance of observations towards fainter extended regions, for those can carry a significant amount of [C\,{\sc ii}] luminosity \citep{Abdullah2020}.

\begin{figure}[tb]
\includegraphics[width=0.5\textwidth, height=0.33\textwidth]{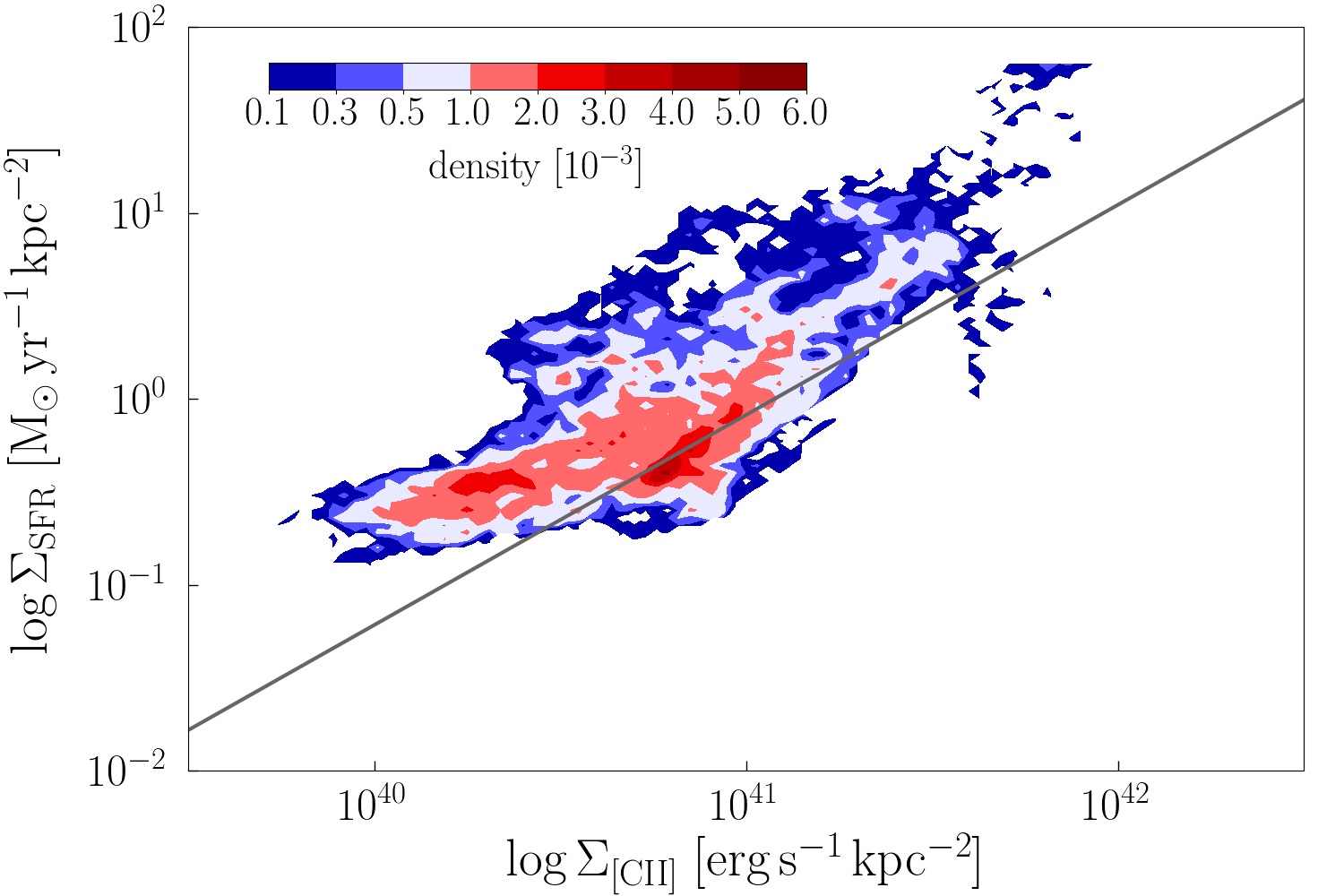}
\caption{SFR surface density, estimated from H$\alpha$ and 24\,$\mu$m emission \citep{Calzetti2007} versus the [C\,{\sc ii}] luminosity per unit area in Orion A as point-density plot. The grey line indicates the relation calibrated by \cite{Herrera-Camus2015} from a sample of local galaxies.}
\label{Fig.SFR-CII}
\end{figure}

\begin{figure}[tb]
\includegraphics[width=0.5\textwidth, height=0.6\textwidth]{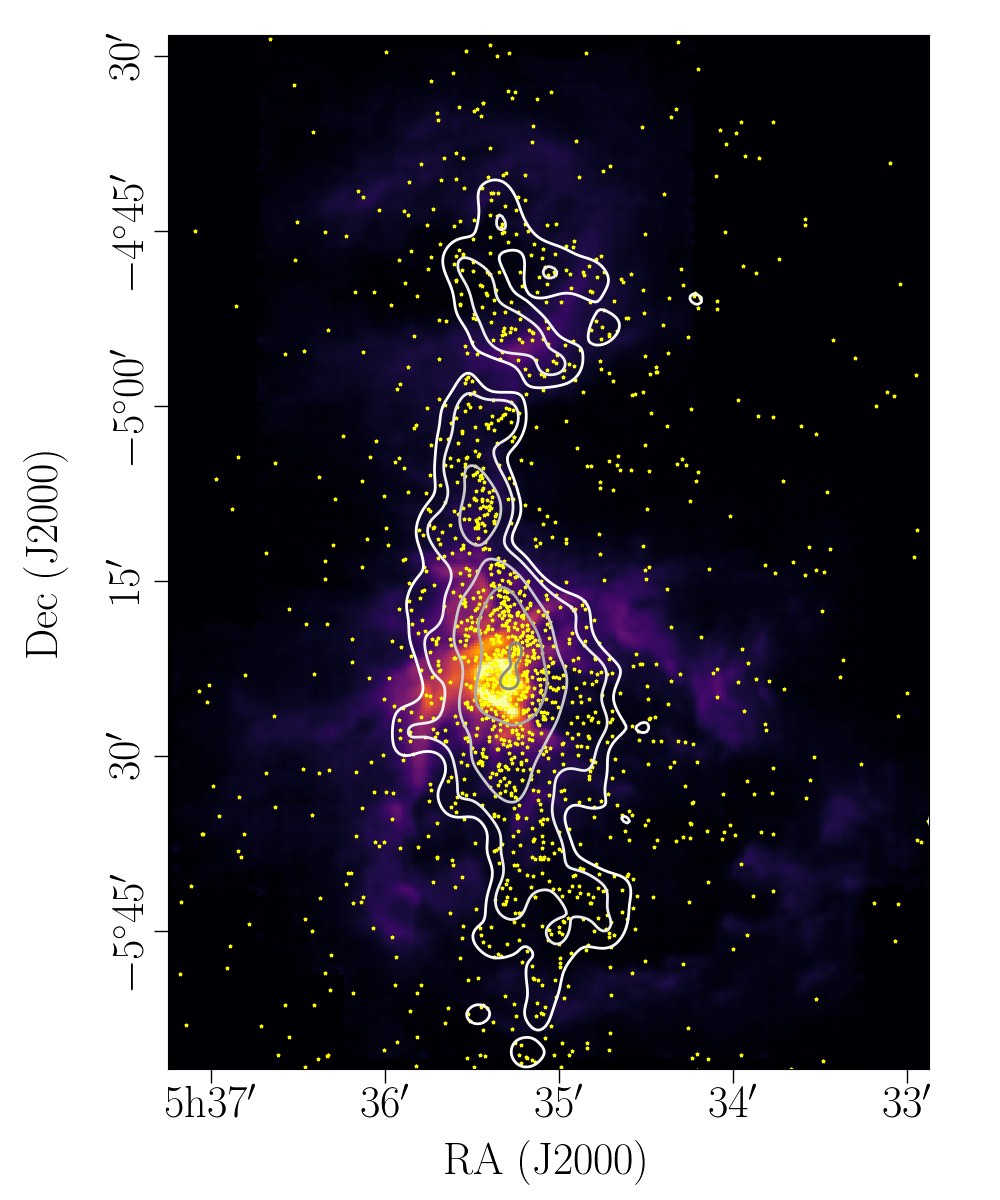}
\caption{[C\,{\sc ii}] line-integrated intensity with YSO surface density in contours (from white to grey: $20, 30, 60, 120, 200\,M_{\sun}\,\mathrm{pc}^{-2}$). Yellow stars mark the positions of the dusty YSOs \citep{Megeath2012}.}
\label{Fig.YSOs}
\end{figure}

\subsection{Tracers of the star-formation rate}

Observations with COBE revealed that the [C\,{\sc ii}] line is the brightest emission line in the far-IR spectrum of the Milky Way \citep{Bennett1994}. As this line originates from gas illuminated by FUV photons emitted by OB stars, it was quickly recognized that this line might provide an excellent probe of the star formation rate in distant galaxies. Indeed, one of the main drivers of the design of ALMA was the requirement to be able to detect this line in Milky Way type galaxies at redshifts of $z\sim 3$. Since then, various Herschel-based studies have calibrated the [C\,{\sc ii}] line as a star-formation rate indicator. We discuss the relations compiled in Table \ref{Tab.SFR_formulae} in the context of our observations towards the Orion Nebula complex\footnote{Since these are empirical relations, we do not correct for [C\,{\sc ii}] optical depth.}. Table \ref{Tab.SFR} summarizes the values we calculate with these formulae for the star-formation rate from observations, both total and divided into the three regions M42, M43 and NGC 1977. 

\begin{table*}[tb]
\centering
\addtolength{\tabcolsep}{12pt}
\def\arraystretch{1.7}
\caption{Star-formation rates from different tracers}
\begin{tabular}{lll}
\hline\hline
Tracer & $\mathrm{SFR(M_{\odot}\,yr^{-1})}$ & Reference \\ \hline
FIR & $4.5\times 10^{-44}\,L({\mathrm{FIR}})(\mathrm{erg\,s^{-1}})$ & \cite{Kennicutt1998} \\
H$\alpha$, $24\,\mu\mathrm{m}$ & $5.3\times 10^{-42}\, (L(\mathrm{H}\alpha)_{\mathrm{obs}} + 0.0031\, L(24\,\mu\mathrm{m}))(\mathrm{erg\,s^{-1}})$ & \cite{Calzetti2007} \\
$24\,\mu\mathrm{m}$ & $1.27\times 10^{-38}\, [L(24\,\mu\mathrm{m})(\mathrm{erg\,s^{-1}})]^{0.8850}$ & \cite{Calzetti2007} \\
H$\alpha$, TIR & $5.5\times 10^{-42}\, [L(\mathrm{H}\alpha)_{\mathrm{obs}} + 0.0024\, L({\mathrm{TIR}})](\mathrm{erg\,s^{-1}})$ & \cite{Kennicutt2009} \\
{[C\,{\sc ii}]}(1) & $5.0\times 10^{-37}\, [L(\mathrm{[C\,\textsc{ii}]})(\mathrm{erg\,s^{-1}})]^{0.89}$ & \cite{Pineda2014} \\
{[C\,{\sc ii}]}(2) & $2.286\times 10^{-43}\, [L(\mathrm{[C\,\textsc{ii}]})(\mathrm{erg\,s^{-1}})]^{1.034}$ & \cite{Herrera-Camus2015} \\
dense gas & $1.5\times 10^{-8}\, M_{\mathrm{dense}}(M_{\odot})$ & \cite{Lada2012} \\ \hline
\end{tabular}
\label{Tab.SFR_formulae}
\end{table*}

\begin{table*}[tb]
\centering
\addtolength{\tabcolsep}{0pt}
\def\arraystretch{1.2}
\caption{Star-formation rates in $M_{\odot}\,\mathrm{yr^{-1}}$ in M42, M43 and NGC 1977 predicted by several tracers using the correlations presented in Table \ref{Tab.SFR_formulae}.}
\begin{tabular}{l|cccccccc}
\hline\hline
Region & FIR & H$\alpha$, $24\,\mu\mathrm{m}$ & $24\,\mu\mathrm{m}$ & H$\alpha$, TIR\tablefootmark{a} & {[C\,{\sc ii}]}(1) & {[C\,{\sc ii}]}(2) & dense gas\tablefootmark{b} & YSOs\\ \hline
all  & $3.0\times 10^{-5}$ & $7.7\times 10^{-5}$ & $1.6\times 10^{-5}$ & $1.0\times 10^{-4}$ & $1.3\times 10^{-4}$ & $1.1\times 10^{-5}$ & $1.5\times 10^{-4}$ & $7.5\times 10^{-4}$\\
M42 & $2.4\times 10^{-5}$ & $7.1\times 10^{-5}$ & $1.2\times 10^{-5}$ & $9.2\times 10^{-5}$ & $9.9\times 10^{-5}$ & $7.6\times 10^{-6}$ & $1.3\times 10^{-4}$ & $5.5\times 10^{-4}$\\
M43 & $2.4\times 10^{-6}$ & $3.9\times 10^{-6}$ & $2.3\times 10^{-6}$ & $5.4\times 10^{-6}$ & $9.9\times 10^{-6}$ & $5.2\times 10^{-7}$ & $3.6\times 10^{-6}$ & --\\
NGC 1977 & $3.1\times 10^{-6}$ & $1.3\times 10^{-6}$ & $3.0\times 10^{-6}$ & $3.1\times 10^{-6}$ & $3.5\times 10^{-5}$ & $2.3\times 10^{-6}$ & $1.4\times 10^{-5}$ & $1.5\times 10^{-4}$\\ \hline
\end{tabular}
\tablefoot{\tablefoottext{a}{Total infrared luminosity approximated by twice the far-infrared luminosity.}
\tablefoottext{b}{Dense gas taken as CO-traced mass.}}
\label{Tab.SFR}
\end{table*}

Perusing Table \ref{Tab.SFR}, we conclude that the different tracers of star-formation predict significantly different values at our mapped spatial scales. Specifically, the two SFR estimates from the [C\,{\sc ii}] luminosity differ from each other by an order of magnitude. Overall, the relation derived by \cite{Herrera-Camus2015} predicts lower values than expected from other traces such as FIR, H$\alpha$ and $24\,\mu\mathrm{m}$ emission. Only in NGC 1977, the relation by \cite{Pineda2014} overpredicts the SFR compared to the other traces (besides CO-traced mass). 

In general, our correlations show that the [C\,{\sc ii}] intensity depends not linearly on the FIR intensity, but by a power law: it increases less steeply than the FIR intensity (Sec. \ref{sec.OrionB}). This reflects the decreased heating efficiency at higher radiation field (Sec. \ref{Sec.pe-efficiency}). This means that the galactic relations of FIR and [C\,{\sc ii}] emission are compatible with each other only within a limited range of conditions. The predictive power of the [C\,{\sc ii}] line in galactic context depends on the dominant source of [C\,{\sc ii}] line emission. A study by \cite{Abdullah2020} of the Orion-Eridanus environment on $\sim 400\,\mathrm{pc}$ scale suggests that [C\,{\sc ii}] emission arises mostly from extended, moderately irradiated cloud surfaces.

We note that the different measures for the star-formation rate trace different evolutionary stages in the star-formation process. While the dense mass mainly traces the sites of potential future star formation, H$\alpha$ traces visible sites of already existing massive stars, that are ionizing their environment. The FIR luminosity also includes contributions from the embedded phase of massive star formation which is missed by H$\alpha$ and $24\,\mu\mathrm{m}$. For galactic sources, this phase is rather short and gets averaged out in the beam, but that is not always the case in extragalactic settings (e.g., ULIRGs with vigorous embedded star formation). Moreover, locally the $24\,\mu\mathrm{m}$ intensity, tracing the warmer shell interiors, is not correlated with the [C\,{\sc ii}], FIR and $8\,\mu\mathrm{m}$ intensities, that trace the somewhat cooler gas in PDR surfaces. Locally, also the ionized gas emitting in H$\alpha$ fills the space interior to the shells.

We compare the gas-derived SFR with an actual count of the YSOs in the Orion Nebula cluster (ONC). \cite{Megeath2016} report 3001 YSOs in the ONC (after sample incompleteness correction). Adopting an average mass of $0.5\,M_{\sun}$ and an average duration of the protostellar/disk phase of $2\,\mathrm{Myr}$ \citep{Evans2009}, gives an SFR of $7.5\times 10^{-4}\,M_{\sun}\,\mathrm{yr^{-1}}$. \cite{DaRio2014} estimate a total mass of the ONC of $1000\,M_{\sun}$. Using the age of the cluster of $2\,\mathrm{Myr}$, this yields a SFR of $5\times 10^{-4}\,M_{\sun}\,\mathrm{yr^{-1}}$. Not surprisingly, as we are zooming in on a region of active star formation, these values are significantly higher than suggested by extragalactic correlations. This is illustrated by comparing the results in Table \ref{Tab.SFR} with the upper-left panel of Figure 2 in \cite{Pineda2018}, the SFR-[C\,{\sc ii}] correlation per unit surface. The Orion Nebula complex\footnote{The total [C\,{\sc ii}] luminosity from the Orion Nebula complex is $750\,L_{\sun}$ from an area of 62\,pc$^2$.} lies off the upper right end of the correlation, with $\log \Sigma_{\mathrm{[C\,\textsc{ii}]}} \mathrm{[erg\,s^{-1}\,kpc^{-2}]}\simeq 40.7$ and $\log \Sigma_{\mathrm{SFR}}\mathrm{[M_{\sun}\,yr^{-1}\,kpc^{-2}]}\simeq 0.9\text{-}1.7$ from star and YSO counts or $\log \Sigma_{\mathrm{SFR}}\simeq 0.1$ from H$\alpha$ and 24\,$\mu$m emission. We note, however, that a pixel-by-pixel comparison of the SFR predicted by H$\alpha$ and 24\,$\mu$m emission \citep[according to][]{Calzetti2007} and the [C\,{\sc ii}] emission in the Orion Nebula, as shown in Fig. \ref{Fig.SFR-CII}, exhibits a similar spread as the SFR-[C\,{\sc ii}] correlation of \cite{Pineda2018} (here, $\rho\simeq 0.78$ and $rms\simeq 0.45\,\mathrm{dex}$), despite the fact that the emission tracers possess different morphologies. In the Orion Nebula, the H$\alpha$ and 24\,$\mu$m emission are strong in the (ionized) shell interior, whereas [C\,{\sc ii}] emission is arising predominantly from the shell's PDR surfaces.

Zooming into separate regions, we discuss the distribution of the YSOs, as catalogued by \cite{Megeath2012}, in relation with [C\,{\sc ii}] emission in somewhat more detail. Fig. \ref{Fig.YSOs} shows the distributions of YSOs across the [C\,{\sc ii}]-mapped area. We have counted the YSOs per $14\arcsec$ pixel and convolved the resulting surface density to a beam of $2\arcmin$. At our spatial resolution, the [C\,{\sc ii}] intensity is not correlated with the YSO surface density locally, which also means a lack of evidence for triggered star formation in this early stage of evolution of the expanding shells in the Orion Nebula complex \citep[see also][]{Goicoechea2020}. Strikingly, while OMC2 does not exhibit bright [C\,{\sc ii}] emission, we find many YSOs (about 120 within $1.6\,\mathrm{pc}^2$) located in the line of sight towards OMC2, concentrated along the ISF. Towards NGC 1977, we find about 300 YSOs across $17\,\mathrm{pc}^2$. The majority of YSOs is located in front of M42 and M43, about 1100 YSOs distributed over $42\,\mathrm{pc}^2$. Here, the YSO surface density is $26\,M_{\sun}\,\mathrm{pc}^{-2}$, where we have used an average mass of $0.5\,M_{\sun}$ and corrected for sample incompleteness by a factor of 2, following \cite{Megeath2016}. The YSO surface density is slightly lower in NGC 1977, $18\,M_{\sun}\,\mathrm{pc}^{-2}$, but highest of the three regions in OMC2 with $75\,M_{\sun}\,\mathrm{pc}^{-2}$. Generally, high YSO surface density is found along the ISF, concentrating in the molecular cores OMC1--4. The maximum surface density is found towards OMC1. Thus, the SFR is highest in M42 and M43, together $6\times 10^{-4}\,M_{\sun}\,\mathrm{yr}^{-1}$. NGC 1977 and OMC2 count with $2\times 10^{-4}\,M_{\sun}\,\mathrm{yr}^{-1}$ and $6\times 10^{-5}\,M_{\sun}\,\mathrm{yr}^{-1}$, respectively. We note that the removal of some 40\% of the mass from the OMC1 core by the stellar wind of $\theta^1$ Ori C will decrease the gravitational binding of the ONC cluster. We expect that it will take some $10^5$ years for the stars to respond.

From COBE observations, \cite{Abdullah2020} derive a [C\,{\sc ii}] flux of $6.9\times 10^{-7}\,\mathrm{erg\,s^{-1}\,cm^{-2}}$ from a region encompassing the Orion Nebula and NGC 1977. This corresponds to $3.7\times 10^3\,L_{\sun}$. Hence, the SFR relations of \cite{Pineda2014} and \cite{Herrera-Camus2015} predict $6\times 10^{-4}\,M_{\sun}\,\mathrm{yr^{-1}}$ and $6\times 10^{-5}\,M_{\sun}\,\mathrm{yr^{-1}}$, respectively. Taking into account the contribution of molecular cloud surfaces exposed to only moderate radiation fields, averaged in the beam of COBE, thus increases the SFR derived from extragalactic correlations. Locally, the SFR can be much higher (like in the Orion Nebula), but on galactic scales larger structures may contribute significantly, as we noted that a large portion of the [C\,{\sc ii}] luminosity from the Orion Nebula complex stems from rather extended, low-intensity regions (cf. Section 4.4).

\section{Conclusion}

Using a square-degree sized map of the [C\,{\sc ii}] $158\,\mu\mathrm{m}$ line towards the Orion Nebula complex, including M43 and NGC 1977, obtained with SOFIA/upGREAT, and comparing with {\it Herschel}/PACS and SPIRE FIR photometry, {\it Spitzer}/IRAC $8\,\mu\mathrm{m}$ emission and velocity-resolved IRAM 30m CO(2-1) observations, we have obtained the following results:

\begin{enumerate}
\item Our comparison of the [C\,{\sc ii}] line emission with FIR emission from FUV-heated dust and PAH emission in the $8\,\mu\mathrm{m}$ band reveal tight correlations with correlation coefficients of $\rho\gtrsim 0.9$. The [C\,{\sc ii}] line-integrated intensity depends less than linear on the PACS $70\,\mu\mathrm{m}$ intensity, FIR ($40\text{-}500\,\mu\mathrm{m}$) intensity, and IRAC $8\,\mu\mathrm{m}$ intensity. In particular, we find that the [C\,{\sc ii}] intensity scales with the $70\,\mu\mathrm{m}$, FIR intensity, and the $8\,\mu\mathrm{m}$ intensity to the power of $\sim 0.6$ each. The FIR and PAH $8\,\mu\mathrm{m}$ intensities are approximately linearly correlated, the IRAC $8\,\mu\mathrm{m}$ band carrying about 10\% of the FIR emission.
\item The less-than-linear dependence of the [C\,{\sc ii}] intensity on the FIR intensity implies a decreasing [C\,{\sc ii}]/FIR ratio with increasing FIR intensity. This is reminiscent of the ``[C\,{\sc ii}] deficit'' in ULIRGs, that has also been observed towards extragalactic sources like M51. However, including [O\,{\sc i}] cooling corrects for most of the deficit in the Orion Nebula complex. The remaining slight deficit can be attributed to a reduced heating efficiency in the heavily irradiated PDR surfaces close to the central stars. Some sight lines host large columns of (warm and cold) dust that produce a ``FIR excess'', leading to particularly low [C\,{\sc ii}]/FIR intensity ratios.
\item Most of the [C\,{\sc ii}] luminosity of the Orion Nebula complex, 33\% of the total $750\,L_{\sun}$, arises in the large, moderately illuminated shells surrounding the Orion Nebula, M43, and NGC 1977. Only 10\% of the [C\,{\sc ii}] luminosity in M42, $53\,L_{\sun}$, stem from the brightest region, OMC1. A large amount of [C\,{\sc ii}] luminosity, 35\%, stems from extended faint surfaces.
\item The scaling relations invoked to estimate the star-formation rate on galactic scales and in high-redshift galaxies do not apply to our local sample of [C\,{\sc ii}]-emitting sources. In the Orion Nebula complex, the [C\,{\sc ii}] line is dominantly excited by FUV radiation from the most massive stars. Smaller sources contribute to small-scale [C\,{\sc ii}]-emitting structures, but minor to the large-scale shells surrounding the massive stars. However, even the total [C\,{\sc ii}] emission from the mapped area is short of predicting the YSO content within the field of view. We surmise that [C\,{\sc ii}] emission from extended [C\,{\sc ii}]-faint structures contribute significantly to [C\,{\sc ii}] emission on galactic scales. In our sample, [C\,{\sc ii}] emission is dominated by larger, comparatively faint structures, rather than the bright dense cores.
\end{enumerate}

Studies of the relations of [C\,{\sc ii}] emission to other tracers of gas and dust help to reveal ``the local truth'' about the origins of the [C\,{\sc ii}] line emission. Further investigations have to show whether the same correlations between those gas and dust tracers hold in different environments, for instance, the starburst region 30 Doradus (with also lower metallicity) and other sources shaped by stellar feedback. Taking into account the geometry of the sources is crucial in understanding the details of the correlations.

\begin{acknowledgements}
This work is based on observations made with the NASA/DLR Stratospheric Observatory for Infrared Astronomy (SOFIA). SOFIA is jointly operated by the Universities Space Research Association, Inc. (USRA), under NASA contract NNA17BF53C, and the Deutsches SOFIA Institut (DSI) under DLR contract 50 OK 0901 to the University of Stuttgart.

This project has received funding from the European Research Council (ERC) under the European Union's Horizon 2020 research and innovation programme (grant agreement No. 851435). JRG thanks the Spanish MCIU for funding support under grants AYA2017-85111-P and PID2019-106110GB-I00. Research on the interstellar medium at Leiden Observatory is supported through a Spinoza award of the Dutch Science Organisation (NWO).

We thank S.~T. Megeath for helpful advice on the {\it Spitzer}/IRAC data.

This work made use of the TOPCAT software \citep{Taylor2005}.
\end{acknowledgements}

\bibliographystyle{aa} 
\bibliography{article3} 

\appendix

\section{Calculation of the ionization parameter $\gamma$}
\label{Sec.gamma}

We use the ionization parameter $\gamma=G_0 T_{\mathrm{gas}}^{1/2}/n_{\mathrm{e}}$ in Section \ref{Sec.pe-efficiency} to semi-empirically determine the heating curve of the interstellar gas. Here, we elaborate the procedure by which we estimate $\gamma$ throughout the [C\,{\sc ii}] map.

To obtain the kinetic temperature of the gas we use the peak temperature $T_{\mathrm{P}}$ of the [C\,{\sc ii}] line:
\begin{align}
T_{\mathrm{gas}} = \frac{T_{\mathrm{ex}}}{1-T_{\mathrm{ex}}/91.2\,\mathrm{K}\log(1+n_{\mathrm{cr}}/n)}
\end{align}
where the critical density depends on the [C\,{\sc ii}] optical depth $\tau_{\mathrm{[C\,\textsc{ii}]}}$, $n_{\mathrm{cr}}=n_{\mathrm{cr,0}}\beta(\tau_{\mathrm{[C\,\textsc{ii}]}})$, and $n_{\mathrm{cr,0}}=2.7\times 10^3\,\mathrm{cm^{-3}}$. The function $\beta(\tau)$ can be approximated as $(1-\exp\tau)/\tau$. For simplicity, we fix $\tau_{\mathrm{[C\,\textsc{ii}]}}=2$, as is consistent with the [$^{13}$C\,{\sc ii}] line analysis where possible (see Paper II). We note that the gas temperature is only weakly dependent on the exact value of $\tau_{\mathrm{[C\,\textsc{ii}]}}$. We estimate the excitation temperature $T_{\mathrm{ex}}$ assuming optically thick emission:
\begin{align}
T_{\mathrm{ex}} = \frac{91.2\,\mathrm{K}}{\log(91.2\,\mathrm{K}/(T_{\mathrm{P}}+J(T_{\mathrm{d}}))+1)}.
\end{align}

We approximate the gas density by the gas column density estimated from the dust opacity, assuming a line of sight of $\sim r/2\simeq 1.4\,\mathrm{pc}$, i.e. $n \sim 6\times 10^{24}\,\mathrm{cm^{-2}}\tau_{160}/1.4\,\mathrm{pc}$ in the Veil shell; in the OMC1 region, we take $n\sim 10^5\,\mathrm{cm^{-2}}$, in accordance with studies of the Orion Bar \citep{Pellegrini2009, Bernard-Salas2012, Goicoechea2015}. We assume that all electrons stem from carbon ionization and that carbon is fully ionized. For the radiation field in the Orion Nebula, we assume $G_0\simeq 500 (1\,\mathrm{pc}/(d+0.1\,\mathrm{pc})^2$, as obtained from the FIR-d relation in Section \ref{Sec.correlations}. We note that for points close to $\theta^1$ Ori C, the estimate of the radiation field using the projected distance is not accurate and we add the vertical offset of $0.1\,\mathrm{pc}$.

\end{document}